\documentclass[10pt]{article}
\usepackage{amsmath,amssymb,tikz}
\usepackage{graphicx}
 \allowdisplaybreaks
\def\be{\begin{equation}}
\def\ee{\end{equation}}
\def\ba#1\ea{\begin{align}#1\end{align}}
\def\bg#1\eg{\begin{gather}#1\end{gather}}
\def\bm#1\em{\begin{multline}#1\end{multline}}
\def\bmd#1\emd{\begin{multlined}#1\end{multlined}}

\def\b{\beta}

\def\e{\epsilon}

\def\r{\rho}
\def\s{\sigma}

\def\la{\label}

\def\nn{\nonumber}

\def\({\left(}
\def\){\right)}
\def\[{\left[}
\def\]{\right]}

\def\Tr{{\rm Tr}}

\def\bea{\begin{eqnarray}}
\def\eea{\end{eqnarray}}
\newcommand{\eq}[1]{(\ref{#1})}
\newcommand{\bit}{\begin{itemize}}  \newcommand{\eit}{\end{itemize}}
\newcommand{\ben}{\begin{enumerate}}  \newcommand{\een}{\end{enumerate}}

\def\la{\langle}
\def\ra{\rangle}
\newcommand{\tr}{\mbox{tr}}

  \def\cC{{\cal C}}

 \def\cK{{\cal K}} 
  
  \def\cR{{\cal R}}

\def\ii{{\rm i}}
\def\ih{{\hat{i}}}
\def\jh{{\hat{j}}}

\long\def\symbolfootnote[#1]#2{\begingroup%
\def\thefootnote{\fnsymbol{footnote}}\footnote[#1]{#2}\endgroup}

\newcommand{\ncts}{{\it Physics Division, National Center for Theoretical Sciences,\\
National Tsing-Hua University, Hsinchu, 30013, Taiwan}}

\newcommand{\nthu}{{\it Department of Physics, National Tsing-Hua
  University, 
Hsinchu 30013, Taiwan}}

\begin{document}
\thispagestyle{empty}
\begin{center}

~\vspace{20pt}

{\Large\bf Universality in the Shape Dependence of Holographic R\'enyi 
Entropy for General Higher Derivative Gravity}

\vspace{25pt}

Chong-Sun Chu\symbolfootnote[1]{Email:~\sf cschu@phys.nthu.edu.tw}
${}^\dagger{}$, 
Rong-Xin Miao\symbolfootnote[2]{Email:~\sf miaorongxin.physics@gmail.com}

\vspace{10pt}${}^\ast{}$\ncts

\vspace{10pt}${}^\dagger{}$\nthu

\vspace{2cm}

\begin{abstract}

We consider higher derivative gravity and obtain universal relations
for the 
shape coefficients  $(f_a, f_b, f_c)$ of the 
shape dependent
universal part of the R\'enyi entropy
for four dimensional CFTs 
in terms of the 
parameters $(c, t_2, t_4)$ of two-point and three-point functions of
stress tensors. As a consistency check, these shape coefficients $f_a$ and $f_c$ 
satisfy the differential relation as derived previously for the
R\'enyi entropy. 
Interestingly, these holographic relations also apply to weakly
coupled 
conformal field theories 
such as theories of free fermions and vectors but are violated by
theories of free scalars. 
The mismatch of $f_a$ for scalars has been observed
in the literature and is due to certain delicate
boundary contributions to the modular Hamiltonian. Interestingly, we
find a combination of our holographic relations which are satisfied
by all free CFTs including scalars. We conjecture that this combined
relation is universal for general CFTs in four dimensional spacetime. 
Finally, we find there are similar universal laws for holographic 
R\'enyi entropy in general dimensions.
 
\end{abstract}

\end{center}

\newpage
\setcounter{footnote}{0}
\setcounter{page}{1}

\tableofcontents

\section{Introduction}

One of the most mysterious features of quantum
mechanics is the phenomena of entanglement. 
For system described by a density matrix $\rho$, entanglement can be
conveniently measured in terms of the entanglement entropy and the 
R\'enyi entropy
\bea
S_{EE} &=& -\Tr(\r \log \r), \\
S_n &=& \frac{1}{1-n} \log \Tr(\r^n).
\eea
For any integer $n>1$, the R\'enyi entropy $S_n$ may be obtained from
\be \label{SZn}
S_n = \frac{\log Z_n -n \log Z_1}{1-n},
\ee
where $Z_n$ is the partition function of the field theory on a certain
$n$-fold branched cover manifold. 
The R\'enyi entropy provides a one parameter family of
entanglememt measurement labeled by an integer $n$, from which  
entanglement entropy $S_{EE}$ can be obtained as a limit 
\be
S_{EE} = \lim_{n\to 1} S_n
\ee
if $S_n$ is continued to
real $n$.

The study of entanglement entropy and the nature of quantum nonlocality
has brought new insights into our understandings of  gravity.
It is found that entanglement plays an important role in the emergence
of space-time and gravitational dynamics
\cite{Ryu:2006bv,
VanRaamsdonk:2010pw,Lashkari:2013koa,Faulkner:2013ica,Lewkowycz:2013nqa}. In
addition to entanglement entropy,
R\'enyi entropy has drawn much attention recently, including the
holographic formula of R\'enyi entropy
\cite{Hung:2011nu,Dong:2016fnf}, the shape dependence of R\'enyi
entropy \cite{Bueno:2015qya,Dong:2016wcf,Dong123}, the holographic
dual of boundary cones \cite{Camps:2016gfs} and R\'enyi twist
displacement operator \cite{Bianchi:2015liz,Balakrishnan:2016ttg}.

Generally, for a spatial region $A$ in a $d$-dimensional spacetime,
the R\'enyi entropy for $A$ is UV divergent.
If one organizes in terms of the short distance cutoff $\e$, one finds
it contain a universal term in the sense that it is independent on the
UV regularization scheme one choose. In odd spacetime 
dimensions, the universal
term is $\e$ independent. In even spacetime dimensions, the universal
term is proportional to $\log \e$ and  its coefficient can be written
in terms of geometric invariant of the entangling surface $\Sigma = \partial
A$. 
In four dimensions, the universal term of the
R\'enyi entropy has the following geometric expansion
\cite{Solodukhin:2008dh,Fursaev:2012mp},
\begin{eqnarray}\label{logterms}
S_n^{\text{univ}}=\log\epsilon\   \left(
\frac{f_a(n)}{2\pi} \cR_\Sigma +
\frac{f_b(n)}{2\pi} \cK_\Sigma -
\frac{f_c(n)}{2\pi} \cC_\Sigma
\right).
\eea
Here the conformal invariants are
\be
\cR_\Sigma \equiv \int_{\Sigma}d^2y\sqrt{\sigma}  R_{\Sigma},\quad
\cC_\Sigma \equiv \int_{\Sigma}d^2y\sqrt{\sigma} C^{ab}_{\ \ ab}, \quad
\cK_\Sigma \equiv \int_{\Sigma}d^2y\sqrt{\sigma} \tr (\bar{K}^2),
\ee
where $\s, R_{\Sigma}, \bar{K}_{\hat{i}\hat{j}}, C^{ab}_{\ \ ab}$ 
are, respectively, the induced metric,
intrinsic Ricci scalar, trace-less part of extrinsic curvature and 
the contraction of the Weyl tensor projected to directions orthogonal 
to the entangling surface $\Sigma$. 
The shape dependence of the R\'enyi entropy is described by 
the coefficients $f_a, f_b, f_c$, which depend on $n$ and the
details of CFTs in general. 
The coefficient $f_a$ can be obtained by studying the thermal free energy of 
CFTs on a hyperboloid \cite{Hung:2011nu}. The coefficients 
$f_c$ and $f_b$ are
determined 
by the stress tensor one-point function and two-point function on the 
hyperboloid background \cite{Bianchi:2015liz,Lewkowycz:2014jia}. 
Remarkably, it is found in  \cite{Lewkowycz:2014jia} that $f_c$ is 
completely determined by $f_a$:
\begin{eqnarray}\label{fafc}
f_c(n)=\frac{n}{n-1}\big{[} f_a(1)-f_a(n)-(n-1)f'_a(n) \big{]}.
\end{eqnarray}
It was conjectured in \cite{Lee:2014xwa} that
\begin{eqnarray}\label{fbfc}
f_b(n)=f_c(n)
\end{eqnarray}
holds for general 4d CFTs. 
This conjecture has passed numerical test for free scalar and
free fermion \cite{Lee:2014xwa}.  
According to \cite{Bianchi:2015liz}, it seems that the relation \eq{fbfc}
holds 
only for free CFTs. Evidence includes an analytic proof for free scalar.
However, it is found to be violated by strongly 
coupled CFTs 
with Einstein gravity duals \cite{Dong:2016wcf}.

In this paper, we apply the holographic approach developed in 
\cite{Dong:2016wcf,Dong123,Balakrishnan:2016ttg} to study 
the
universal terms of the R\'enyi entropy for 
CFTs in general spacetime dimensions
that admit general higher derivative gravity duals. For 4d CFTs,
expanding the coefficients $(f_a, f_b, f_c)$ in powers of 
$(n-1)$, we find the
leading and sub-leading terms are related to  parameters $(c, t_2,
t_4)$ of two point and three 
point functions of stress tensors \cite{Hofman:2008ar,Buchel:2009sk}:
\begin{eqnarray}\label{mainresultfa}
&&f_a(n)=a -\frac{c}{2}(n-1)+
  c(\frac{35}{54}+\frac{7}{324}t_2+\frac{1}{81}t_4)
(n-1)^2+O(n-1)^3\\\label{mainresultfb}
&&f_b(n)=c - c(\frac{11}{12}+\frac{1}{18}t_2
+\frac{1}{45}t_4)(n-1)+O(n-1)^2\\\label{mainresultfc}
&&f_c(n)=c - c(\frac{17}{18}+\frac{7}{108}t_2
+\frac{1}{27}t_4)(n-1)+O(n-1)^2.
\end{eqnarray}
It should be mentioned that the expansion (\ref{mainresultfa}) of
$f_a$ has been obtained in \cite{Lee:2014zaa} by using two-point and
three-point function of the modular Hamiltonian. Here we provide a
holographic proof of it.
We note that  \eq{mainresultfa} and \eq{mainresultfc}
satisfy the relation \eq{fafc}. This can be regarded as a 
check of our holographic calculations. 
We also note that $t_2=t_4=0$ for Einstein gravity and the
eqs.(\ref{mainresultfb},\ref{mainresultfc}) 
reduce to the results obtained in  \cite{Dong:2016wcf} in this case. 
To the best of our knowledge,
the universal dependence of $f_b$ on the
coefficients $t_2, t_4$ as obtained in 
the relation \eq{mainresultfb}  
is  new. This is one of the main results of this
paper.

We remark that our holographic relations
eqs.(\ref{mainresultfa}-\ref{mainresultfc}) are also satisfied by free
fermions and vectors \footnote{ We have assume $f_b''(1)=f_c''(1)$ for  free
fermions and vectors. Numerical calculations support this assumption for free fermions \cite{Lee:2014xwa}}. However, mismatch appears for free
scalars. Actually, the discrepancy of $f_a$ in scalars has been
observed in \cite{Lee:2014zaa}, which is due to the boundary
contributions to the modular Hamiltonian. 
It was found that the boundary terms in the
stress tensor of scalars are important at weak  coupling and are
suppressed in the strong coupling limit \cite{Lee:2014zaa}. Although
eqs.(\ref{mainresultfa},\ref{mainresultfb},\ref{mainresultfc}) are 
not satisfied by theories of free scalars, 
we find 
that the following combinations 
\begin{eqnarray}
\label{Conjecture}
&&2f'_b(1)-3f'_c(1)=c(1+\frac{1}{12}t_2+\frac{1}{15}t_4),\\ 
\label{Conjecture1}
  &&2f'_b(1)+\frac{9}{2}f''_a(1)=c(4+\frac{1}{12}t_2+\frac{1}{15}t_4),
\end{eqnarray}
are satisfied by all CFTs with holographic dual and all free CFTs
including free scalars. We conjecture they are universal 
relations for all
CFTs in four dimensions. Note that we have
$f'_c(1)+\frac{3}{2}f''_a(1)=c$ from eq.(\ref{fafc}), therefore
eq.(\ref{Conjecture}) and eq.(\ref{Conjecture1}) are not
independent. Without loss of generality, we focus on the conjecture
eq.(\ref{Conjecture}) in the rest of this paper.

In the notation of \cite{Bianchi:2015liz}, 
our conjecture (\ref{Conjecture}) 
for 4d CFTs 
can be written in the form
\be\label{ConjectureNew}
\pi  C_D''(1)-36 h_n''(1)
=\frac{2 \pi^3}{5} C_T(1+\frac{1}{12}t_2+\frac{1}{15}t_4), 
\ee
where $C_T=\frac{40}{\pi^4}c$, 
$h_n(n)$ and $C_D(n)$ are CFT data associated with the presence of the
entangling surface.
In general, for a $d$-dimensional CFT and an  entangling surface
$\Sigma$ (codimension 2),
one denotes the coordinates  orthogonal and
parallel to the entangling surface by $x^a$ and $y^\ih$.
The breaking
of translational invariance in the directions transverse to $\Sigma$ 
can be characterized by the displacement operators $D^a (y^\ih)$.
As a result, one has the following correlation functions 
\cite{Bianchi:2015liz}:
\begin{eqnarray}\label{hn}
&&\la T_{\ih \jh}  \ra_n
=-\frac{h_n(n)}{2\pi n}\frac{\delta^{\ih \jh}}{|x^a|^d} ,\\ 
&&\la D^a(y^\ih)D^b(0)\ra_n=C_D(n) \frac{\delta^{ab}}{|y^\ih|^{2(d-1)}}.
\end{eqnarray}
Here
$h_n(n)$ is 
the coefficient fixing the normalization of 
the one-point function for the stress tensor in the presence of the
twisted operator for the $n$-fold replicated QFT,
and  $C_D(n)$ is the normalization 
coefficient for the two-points correlation function of 
the displacement operators. 
In 4-dimensions, 
$C_D(n)$ and $h_n(n)$ related to the dependence 
of R\'enyi and entanglement entropy
on smooth or shape deformations 
\cite{Bueno:2015qya,Bueno:2015rda,Mezei:2014zla,Faulkner:2015csl}. 
The specific relation can be found in 
eqs.(2.12, 3.15, 3.19) of \cite{Bianchi:2015liz}.

It should be mentioned that 
unlike  $f_c$ and $f_b$ which are defined only in 4 dimensions, 
$h_n$ and $C_D$ have a natural definition in all dimensions. 
Therefore it is natural to 
ask if by using them one can generalize
the results \eq{Conjecture} and \eq{Conjecture1} 
to other dimensional spacetime. The holographic dual of $h_n$ and
$C_D$ for Einstein gravity and Gauss-Bonnet Gravity in general
dimensions are studied in recent works
\cite{Balakrishnan:2016ttg,Dong123}. Applying their results, one can
express $h_n''(1)$ and $C_D''(1)$ in terms of $C_T$ and $t_2$. Recall
that we have $t_4=0$ for Einstein gravity and Gauss-Bonnet Gravity. To
get the information of $t_4$, one has to study at least one cubic
curvature term such as $\mathcal{K}_7$ and  $\mathcal{K}_8$ in the
action (\ref{Kni}).  Following the approach of \cite{Dong123, 
Balakrishnan:2016ttg}, we obtain the holographic 
formulae of 
$h_n$ and $C_D$ for
a $d$-dimensional CFT admiting a general higher curvature gravity dual: 
\begin{eqnarray}\label{hnNEW}
&&\frac{h_n}{ C_T}=-2\pi n \frac{ M_e}{f_d},\\ \label{GHNEW}
&& \frac{C_D}{C_T}= \frac{d \pi ^2 n}{d+1}\big{[} (d-2) 
(\beta_{n}-\beta_{1})-\frac{M_e}{2} \big{]},
\end{eqnarray}
where $M_e$ is the effective mass defined in eq.(\ref{GHCfF})
and $\beta_n$ is the coefficient in the function $k(r)$
in eq.\eq{GHCk} which describes a deformation in the extrinsic curvature of
the entangling surface. 
It is remarkable that
these relations take simple and universal form for all the higher 
curvature gravity.

By using the holographic formula of $h_n$ and $C_D$, we find there are 
similar universal laws in general dimensions, 
which 
involves linear combinations of the terms 
$C_D''(1), h_n''(1), C_T, C_T t_2$ and $C_T t_4$ 
\footnote{ In three dimensions, we have $t_2$=0. And  we have
  $t_2=t_4=0$ in two dimensions.}. 
In general, we have
for a $d$-dimensional CFT, 
\begin{eqnarray}\label{secondDhnnew}
\frac{h_n''(1)}{C_T}&=&-\frac{2 \pi ^{\frac{d}{2}+1} \Gamma
  \left(\frac{d}{2}\right)}{(d-1)^3 d (d+1) \Gamma (d+3)} \big{[} d
  \left(2 d^5-9 d^3+2 d^2+7 d-2\right)\nonumber\\
&+&(d-2) (d-3) (d+1) (d+2) (2 d-1) t_2+(d-2) \left(7 d^3-19 d^2-8
  d+8\right)t_4 \big{]}\\ \label{secondDCDnew}
\frac{C_D''(1)}{C_T}&=&\frac{4 \pi ^2}{d+1}\big{[}
  \frac{1-d^2+d}{d^2-d}-\frac{(d-2)(d-3) }{(d-1)^2 d}\  t_2
  -\frac{(d-2) \left(3 d^2-7 d-8\right)}{(d-1)^2 d (d+1) (d+2)}\  t_4
  \big{]}.
\end{eqnarray}
Note that the relation
 $C_D''(1)=d\Gamma(\frac{d+1}{2})(\frac{2}{\sqrt{\pi}})^{d-1}h_n''(1)$
is obeyed by free fermions and conformal tensor fields
\footnote{The conformal tensor fields appear only in even dimensions.}
but are violated by free scalars.
However similar to the 4 dimensional case, 
there exist  'universal laws' that include free scalar fields. 
For example, in three dimensions, we find
\begin{eqnarray}\label{3duniversallawNew}
\pi  C_D''(1)-16 h_n''(1)=\frac{ \pi ^3}{3} C_T(1+\frac{t_4}{30}),
\end{eqnarray}
works well for free fermions, free scalars and CFTs with gravity dual. As for the 'universal laws' in higher dimensions,  please refer to eq.(\ref{universallawinGD}). It is interesting to study whether these 'universal laws' are obeyed
by more general CFTs.

The paper is organized as follows. In Sect.2, we study 
4d CFTs which are dual
to general higher curvature gravity and derive the relations between
the coefficients $(f''_a(1), f'_b(1), f'_c(1))$ in the universal terms
of R\'enyi entropy and the parameters $(c, t_2, t_4)$ of two point and
three point functions of  the stress tensors in the conformal field
theory.  In Sect. 3, we compare these holographic relations with those
of free CFTs and find a combined relation which agrees with all the
known results of the free CFTs. We conjecture this combined relation
is a universal law for all the CFTs in four dimensions. In Sect. 4, we
consider three and higher general spacetime dimensions and 
derive the holographic dual of $h_n$ and $C_D$ for general higher
curvature gravity and discuss the universal behaves of  $h_n''(1)$ and 
$C_D''(1)$.  Finally, we
conclude in Sect.5.

Notations: We use $x^{\mu}$ ($y^i$) and $g_{\mu\nu}$ ($\gamma_{ij}$)
to denote the coordinates and metric in the bulk (on the
boundary). $x^a$ and $y^{\hat{i}}$ are the orthogonal and parallel
coordinates on the entangling surface. $\sigma_{\hat{i}\hat{j}}$ is the
induced metric on the entangling surface. For simplicity, we focus on  Euclidean signature in this paper.

\section{Holographic R\'enyi Entropy for Higher Derivative Gravity}

In this section, we investigate the universal terms of R\'enyi entropy
for 
4d CFTs 
that are dual to general higher derivative gravity. We firstly take 
Gauss-Bonnet gravity as an example and then generalize the results to 
general higher curvature gravity. Some interesting relations  between
the 
universal terms of 
holographic R\'enyi entropy (HRE) 
and the parameters of two point and
three point functions of stress tensors are found. 

\subsection{Gauss-Bonnet Gravity}

For simplicity, we consider the following Gauss-Bonnet Gravity which
is 
slightly different from the standard form
\begin{eqnarray}\label{GBaction}
I=\frac{1}{16\pi G_N}\int_{M} \big{[} R+\frac{12}{l^2}
+\alpha
(\bar{R}_{\mu\nu\rho\sigma}\bar{R}^{\mu\nu\rho\sigma}-4\bar{R}_{\mu\nu}
\bar{R}^{\mu\nu}+\bar{R}^2) \big{]}+I_B, 
\end{eqnarray}
where  
$\int_M \equiv \int_{M}d^{d+1}x\sqrt{g}$ ($d=4$ here),
the quantities $\bar{R}_{\cdots}^{\cdots}$ are given by
\begin{eqnarray}\label{barR}
&&\bar{R}_{\mu\nu\rho\sigma}=R_{\mu\nu\rho\sigma}-\frac{1}{l^2}(
g_{\mu\sigma}g_{\nu\rho}-g_{\mu\rho}g_{\nu\sigma}),\\
&& \bar{R}_{\mu\nu}=R_{\mu\nu}+\frac{4}{l^2}g_{\mu\nu},\\
&&\bar{R}=R+\frac{20}{l^2}
\end{eqnarray}
and  $I_B$ denotes the Gibbons-Hawking-York terms which make a 
well-defined variational principle and the counter terms  which make
the total action finite. 

An advantage of the above action is that, similar to Einstein gravity, 
the radius of AdS is exactly $l$. While in the standard GB and higher 
derivative gravity, the effective radius of AdS is a complicated
function 
of $l$, which makes the calculations complicated. Below we set $l=1$
for  simplicity. 

\subsubsection{$f_a(n)$ }

Let us briefly review the method to derive $f_a(n)$
\cite{Hung:2011nu}.  We focus on the spherical entangling surface,
where $\tr \bar{K}^2 $ and $C^{ab}_{\ \ ab}$ vanish. Thus only $f_a$
appears in the universal terms  of R\'enyi entropy
eq.(\ref{logterms}). The main idea is to map the vacuum  state of the
CFTs in a spherical entangling region to the thermal state of CFTs on
a hyperboloid. The later has a natural holographic dual in the bulk,
the black hole that asymptotes to the hyperboloid. Using the free
energy of black hole, we can derive R\'enyi entropy as
\begin{eqnarray}\label{Renyi0}
S_n=\frac{n}{1-n}\frac{1}{T_0}[ F(T_0)-F(T_0/n)]
\end{eqnarray}
where $T_0$ is the temperature of hyperbolic black hole 
for $n=1$. Further using the thermodynamic identity, $S=-\partial F/
\partial T$, 
we can rewrite the above expression as
\begin{eqnarray}\label{GBRenyi}
S_n=\frac{n}{n-1}\frac{1}{T_0}\int_{T_0/n}^{T_0}S_{\text{BH}}(T)dT
\end{eqnarray}
where $S_{\text{BH}}(T)$ is the black hole entropy. 
For our revised GB gravity (\ref{GBaction}), it takes the form
\begin{eqnarray}\label{GBBHentropy}
S_{\text{BH}}=\frac{1}{4G_N}\int_H dy^3\sqrt{h}[1+ 2\alpha (\mathcal{R}_{H}+6)]
\end{eqnarray}
where $H$ denotes horizon and $\mathcal{R}_{H}$ is the intrinsic Ricci
scalar on 
horizon. 

The key point in this approach is finding the black hole solution that
asymptotes to the hyperboloid on the boundary. We get
\begin{eqnarray}\label{GBmetric}
ds^2_{\text{bulk}}=\frac{dr^2}{f(r)}+f(r)d\tau^2+r^2 d\Sigma^2_3
\end{eqnarray}
where $ d\Sigma^2_3$ is the line element for hyperbolic plane $H^3$
with unit 
curvature, and $f(r)$ is given by
\be\label{GBf}
f(r)=\frac{(1+12 \alpha) r^2-\sqrt{8 \alpha  M+(1+8 \alpha)^2
      r^4}}{4 \alpha }-1.
\ee
Here 
\be
M=\left(r_H^2-1\right) \left(( 1+10 \alpha) r_H^2-2 \alpha \right),
\ee
and $r_H$ denotes the position of horizon, $f(r_H)=0$.
Note that $f(r)$ has the correct limit: it becomes that of hyperbolic black hole (black hole in Einstein gravity) when $M\to 0$ ($\alpha\to 0$). In the large $r$ limit, the boundary metric is conformal equivalent to 
\begin{eqnarray}\label{GBboundarymetric}
ds^2_{\text{boun}}=d\tau^2+ d\Sigma^2_3,
\end{eqnarray}
which is the expected metric on manifold $S^1\times  H^3$.

To determine $r_H$, we note that 
the Hawking temperature on horizon is given by
\begin{eqnarray}\label{Hawkingtem0}
T=\frac{1}{4\pi}\partial_r f(r)|_{r=r_H}=\frac{1}{2\pi n}.
\end{eqnarray}
From the above equation, one can easily get $T_0=\frac{1}{2\pi}$ for 
the hyperbolic black hole with $f(r)=r^2-1$ and $r_H=1$. 
Now let us solve eq.(\ref{Hawkingtem0}) and express $r_H$ in terms of $(n-1)$
\begin{eqnarray}\label{GBrh}
r_H=1-\frac{n-1}{3}+\frac{(10+96 \alpha) }{27(1+8 \alpha
  )} (n-1)^2 -\frac{2 \left(49+912 \alpha+
4224 \alpha ^2\right)} {243 (1+8 \alpha)^2}(n-1)^3 +O(n-1)^4.
\end{eqnarray}
Substituting eq.(\ref{GBrh}) together with $T=\frac{1}{2\pi n},
T_0=\frac{1}{2\pi}$ and $\mathcal{R}_{H}=-6/r_H^2$ into
eqs.(\ref{GBRenyi},\ref{GBBHentropy}), 
we obtain
\begin{eqnarray}\label{GBRenyi1}
S_n = \frac{V_{\Sigma}}{4G_N} \left[  1-\frac{1+8\alpha}{2} (n-1)
+\frac{7}{54} (5+48 \alpha) (n-1)^2-\frac{\left(11856 \alpha ^2+2514 
\alpha +133\right) }{162 (8 \alpha +1)} (n-1)^3 \right]\nonumber\\
 +O(n-1)^4,  \; 
\end{eqnarray}
where $V_{\Sigma}$ is the hyperbolic volume, which 
contributes a logarithmic term 
$V_{\Sigma}^{\rm univ}=2\pi \log \epsilon$ 
\cite{Hung:2011nu}. Now we can extract $f_a$ from eq.(\ref{GBRenyi1}) as
\begin{eqnarray}\label{GBfa}
f_a(n)&=&a-\frac{c}{2} (n-1)+\frac{7}{54} (6c-a) (n-1)^2+ O(n-1)^3\nonumber\\
          &=&a -\frac{c}{2}(n-1)+ c(\frac{35}{54}+
\frac{7}{324}t_2)(n-1)^2+O(n-1)^3,
\end{eqnarray}
where $a=\frac{\pi}{8G_N}$ and $c=\frac{\pi}{8G_N}(1+8\alpha)$ 
\cite{Miao:2013nfa}. In the above derivation we have used 
\be 
\frac{c-a}{c}=\frac{1}{6}t_2+\frac{4}{45}t_4 , \quad 
\mbox{for 4d CFTs}
\ee
and $t_4=0$ for GB gravity. Clearly, eq.(\ref{GBfa}) agrees with
eq.(\ref{mainresultfa}) when $t_4=0$. To get the information of $t_4$,
one must consider more general higher derivative gravity. We leave
this problem to next section. Notice that the $O(n-1)^3$ terms of
$f_a$  eq.(\ref{GBRenyi1}) is a complicated function of $a$ and $c$,
which implies  that there is no universal relations at this and higher
orders. From the viewpoint of CFTs, terms of $f_a$ at order $O(n-1)^3$
are determined by four-point functions of stress tensor
\cite{Lee:2014zaa}. Unlike two-point and three-point functions,
four-point functions of CFTs are no longer universal. Thus, it is
expected that there is no universal relation at order $O(n-1)^3$  for
$f_a$. It depends on the details of CFTs at this and higher orders.
Similarly, one expects there is no universal relation at order
$O(n-1)^2$ for $f_b$ and $f_c$.

\subsubsection{$f_c(n)$}

Now let us continue to derive $f_c(n)$. We take the approach developed
in \cite{Dong:2016wcf}.
In general with a deformation of the field theory metric,
the change in the partition function is govern by 
one-point function
of the field theory stress tensor
\begin{eqnarray}\label{partitionfirstorder}
\delta \log Z_n=\frac{1}{2}\int_{\partial M} dx^4
\sqrt{\gamma} \; 
\la T^{ij}\ra 
\delta \gamma_{ij}.
\end{eqnarray}
The main idea of \cite{Dong:2016wcf} is to  consider
specific deformation of the metric so that, on using \eq{SZn}, 
one may isolate the required shape dependent term in the universal
part of the R\'enyi entropy. 
For example, $f_c$ can be isolated with a deformation that affects
$\cC_\Sigma$ but not $\cK_\Sigma$:
\begin{eqnarray}\label{Renyifirstorder}
\delta S_n= -\log\epsilon\   \int_{\Sigma}d^2y\sqrt{\sigma} 
\frac{ f_c(n)}{2\pi}C^{ab}_{\ \ ab}+ \cdots ,
\end{eqnarray}
where $\cdots$ are non-universal terms of the R\'enyi entropy. 
This can be achieved by considering on the entangling surface
the following  metric
\begin{eqnarray}\label{deformedmetricQ}
ds^2_{\text{boun}}=d\tau^2+ \frac{1}{\rho^2}
\big{[}d\rho^2+(\delta_{\hat{i}\hat{j}}
+Q_{ab\hat{i}\hat{j}}x^ax^b+O(\rho^3))dy^{\hat{i}} dy^{\hat{j}} \big{]}
\end{eqnarray}
where $Q_{ab\hat{i}\hat{j}}$ describes a 
deformation of the metric and give rises to an amount of $C^{ab}{}_{ab}$ as
\be
C^{ab}{}_{ab} = \frac{1}{3} Q_a{}^{a\hat{i}}{}_{\hat{i}}.
\ee
Here in \eq{deformedmetricQ}, we have adopted a local 
coordinate system $(\rho, \tau, y^{\hat{i}})$ near $\Sigma$, where
for each point on $\Sigma$, we introduce a  
one-parameter family of geodesics orthogonal to $\Sigma$ 
parametrized by $\tau$, and $\rho$ denotes the radial distance to $\Sigma$ along
such a geodesic.  $(x^1, x^2) \equiv (\rho \cos \tau, \rho \sin \tau)$
and $\{ y^{\hat{i}}, i= 1, \cdots, d-2 \}$ 
denotes an arbitrary coordinates system on $\Sigma$.
We  note that in this computation of $f_c$, the
boundary metric  \eq{deformedmetricQ} is conformal equivalent to
a deformed conical metric.

To proceed with the calculation of $f_c(n)$, 
we consider the bulk metric that asymptotes to the
deformed hyperboloid background \eq{deformedmetricQ}:
\begin{eqnarray}\label{GBmetricQ}
ds^2_{\text{bulk}}=\frac{dr^2}{f(r)}+f(r)d\tau^2+\frac{r^2}{\rho^2}
\big{[}d\rho^2+(\delta_{\hat{i}\hat{j}}+q(r)Q_{ab\hat{i}\hat{j}}x^ax^b+O(\rho^3))
dy^\ih dy^\jh \big{]}
\end{eqnarray}
where $q(r)$ is determined by the E.O.M in the bulk and approach 
1 in the limit $r\to \infty$. Actually, to derive $f_c(n)$, we 
do not need to solve the E.O.M. That is because we already have 
$\delta\gamma_{ij}=\frac{r^2}{\rho^2}Q_{ab\hat{i}\hat{j}}x^ax^b \sim
C^{ab}_{\ \ ab}$, 
so we only need zero order 
of $T^{ij}$  in eq.(\ref{partitionfirstorder}) in order to extract 
the terms proportional to $ C^{ab}_{\ \ ab}$. In other words, we only 
need to calculate $T^{ij}$ on undeformed 
hyperboloid background.

We note that in the context of AdS/CFT,
the stress tensor that appears in \eq{partitionfirstorder}
can be taken  either as the regularized Brown-York boundary stress
tensor \cite{Balasubramanian:1999re} 
or the holographic stress tensor \cite{deHaro:2000vlm}. 
The two are
equivalent as we demonstrate in the appendix. 
In this section, we will consider the first approach.
The key point is to find the 
regularized boundary stress tensor for our non-standard GB gravity 
(\ref{GBaction}). Notice that our non-standard GB gravity
(\ref{GBaction}) 
can be rewritten into the standard form, with only the coefficients of 
$\mathcal{L}_0=1$ and $\mathcal{L}_2=R$ different from the standard GB:
\begin{eqnarray}\label{rewriteGB}
I&=&\frac{1}{16\pi G_N}\int_M \left[
  R+\frac{d^2-d}{l^2}
+\alpha \mathcal{L}_4(\bar{R}) \right], \nonumber\\
&=&\frac{1}{16\pi G_N}\int_{M}
  \left[ \big{(}1+2(d-1)(d-2)
\alpha\big{)}R+\frac{d^2-d}{l^2}\big{(}1+(d+1)(d-2)\alpha\big{)}+
\alpha\mathcal{L}_4(R) \right],
\end{eqnarray}
where 
$\mathcal{L}_4(R)$ denotes the standard GB term. 
The holographic regularization for GB gravity is studied in 
\cite{Brihaye:2008xu}.  Reparameterizing their formulas, we get
for the Brown-York boundary stress tensor:
\begin{eqnarray}\label{GBTij}
8\pi G_NT_{\partial M}^{ij}&=&\big{(}1+2\alpha(d-1)(d-2)\big{)} 
\left[ K_{\partial M}^{ij}-K_{\partial M}
  \gamma^{ij}-(d-1) \gamma^{ij}+
\frac{\Theta(d-3)}{d-2} (R_{\partial M}^{ij}
-\frac{1}{2}R_{\partial M} \gamma^{ij})  \right] \nonumber\\
&&+2\alpha (Q^{ij}-\frac{1}{3}Q \gamma^{ij}),
\end{eqnarray}
where $\Theta(x)$ is the step-function with $\Theta(x)=1$ provided
$x\ge 0$, 
and zero otherwise. $K_{\partial M}^{ij}$ is the extrinsic curvature on the AdS boundary and $Q_{ij}$ is given by
\begin{eqnarray}\label{GBQij}
Q_{ij} &=& 2K_{\partial M} K_{\partial M ik}K_{\partial M\  j}^{\ \ \ \  k}-2K_{\partial M ik} K_{\partial M}^{kl}K_{\partial M lj}+ K_{\partial M ij}(
K_{\partial M kl}K_{\partial M}^{kl}-K_{\partial M}^2)
 \\
&&+2K_{\partial M} R_{\partial M ij}+ R_{\partial M}K_{\partial M ij}-2K_{\partial M}^{kl} R_{\partial M kilj}-4  R_{\partial M ik}K_{\partial M \ j}^{\ \ \ k}. \nn
\end{eqnarray}
Here $R_{\partial M}$ denotes the intrinsic curvature on the boundary.

Substituting eq.(\ref{GBTij}) and  
$\delta\gamma_{ij} =\frac{r^2}{\rho^2}Q_{ab\hat{i}\hat{j}}x^ax^b$ 
into eqs.(\ref{partitionfirstorder},\ref{Renyifirstorder}), we obtain
\begin{eqnarray}\label{GBfc}
f_c(n)&=&\frac{\pi}{8G_N}\left[
1+8 \alpha+\left(-\frac{17}{18}-\frac{32 \alpha }{3}\right) (n-1) 
+\frac{217+4512 \alpha+23232 \alpha ^2 }{162(1+8\alpha)}(n-1)^2\right]
+ O(n-1)^3\nonumber\\
&=&c+(\frac{7}{18}a-\frac{4}{3}c)(n-1)+O(n-1)^2\nonumber\\
&=&c+c(-\frac{17}{18}-\frac{7}{108}t_2)(n-1)+O(n-1)^2
\end{eqnarray}
Similar to $f_a(n)$, we have used
$\frac{c-a}{c}=\frac{1}{6}t_2+\frac{4}{45}t_4$ 
for 4d CFTs and $t_4=0$ for GB gravity. 
Eq.(\ref{GBfc}) agrees with eq.(\ref{mainresultfc}) when $t_4=0$.
Note that eq.(\ref{GBfc}) and eq.(\ref{GBfa}) are consistent with the
identity (\ref{fafc}). This can be regarded as a check of our
holographic calculations.

\subsubsection{$f_b(n)$}

Now let us go on to calculate $f_b(n)$. The method is 
similar to
that  of $f_c(n)$: we consider the first order variation 
\eq{partitionfirstorder} of 
the partition on the hyperboloid background deformed by an 
extrinsic curvature  \cite{Dong:2016wcf}
and then extract  $f_b(n)$ from 
\begin{eqnarray}\label{Renyifirstorderfb}
\delta S_n
=\log\epsilon\   \int_{\Sigma}d^2y\sqrt{\sigma}\frac{f_b(n)}{2\pi}
\text{tr}  (\bar{K}^2)  + \cdots . 
\end{eqnarray}
The main difference from $f_c(n)$ is that now we need to calculate
$T^{ij}$ on 
the deformed hyperboloid $\tilde{H^4_n}$. 
This is because we have $\delta \gamma_{ij}\sim K$, thus 
to extract $K^2$ terms, we must get $T^{ij}$  of order $K$. 

To proceed, we deform the boundary hyperboloid by a traceless  
extrinsic curvature
\begin{eqnarray}\label{deformedmetricK}
ds^2_{\text{boun}}=d\tau^2+ \frac{1}{\rho^2} \big{[}d\rho^2+
(\delta_{\hat{i}\hat{j}}+K_{a\hat{i}\hat{j}}x^a+O(\rho^2))dy^{\hat{i}} dy^{\hat{j}} \big{]}.
\end{eqnarray}
Then the bulk metric becomes
\begin{eqnarray}\label{GBmetricK}
ds^2_{\text{bulk}}=\frac{dr^2}{f(r)}+f(r)d\tau^2+\frac{r^2}{\rho^2}
\big{[}d\rho^2+(\delta_{\hat{i}\hat{j}}+k(r)K_{a\hat{i}\hat{j}}x^a+O(\rho^2) )dy^\ih dy^\jh \big{]}
\end{eqnarray}
To get boundary stress tensor $T^{ij}$  of order $O(K)$, we need to 
solve the E.O.M up to $O(K)$. 
For traceless $K_{aij}$, we find one independent equation 
\begin{eqnarray}\label{GBequationk}
&&\left[
-2 \alpha  \left(f f''+r f'-6 \left(f+r^2\right)\right)+f+r^2\right] 
k(r)\nonumber\\
&&+f \left[ 
r f  \left(-36 \alpha +2 \alpha  f''-3\right)-f' \left((12 \alpha +1)
r^2-4 \alpha  f\right)+2 \alpha  r \left(f'\right)^2\right]  k'(r)\nonumber\\
&&-r f^2  \left(-2 \alpha  f'+12 \alpha  r+r\right) k''(r)=0
\end{eqnarray}
Near the horizon, the solutions behave like $k(r)\sim (r-r_H)^{n/2}$.
The solution is uniquely determined by this IR boundary boundary
condition and the UV boundary condition $\lim_{r\to
  \infty}k(r)=1$. However, the IR boundary boundary condition
$k(r)\sim (r-r_H)^{ n/2}$ is not easy to deal with. Thus, we define a
new function  
\begin{eqnarray}\label{GBh}
h(r)=k(r) \exp\left[\int_r^{\infty}\frac{dr}{f(r)}\right]
\end{eqnarray}
and the E.O.M becomes
\begin{eqnarray}\label{GBequationh}
&&\left[
2 \alpha  (r-1) f''+4 \alpha  f'+(12 \alpha +1) (-(3 r-1))\right] 
h(r)\nonumber\\
&&+\left[f \left(r \left(-36 \alpha +2 \alpha  f''-3\right)
+4 \alpha  f'\right)-r \left(f'+2\right) \left(-2 \alpha  f'+12 \alpha
r+r\right)\right]h'(r)\nonumber\\ 
&&+\left[-f r \left(-2 \alpha  f'+12 \alpha  r+r\right)\right] h''(r)=0
\end{eqnarray}
Now the regularity condition at horizon simply requires $h(r_H)$ to be
finite. Solving the above equation perturbatively, we get
\begin{eqnarray}\label{GBhsolution}
h(r)=\frac{r+1}{r}+h_1(r) (n-1) +h_2(r) (n-1)^2 + \cdots
\end{eqnarray}
with
\begin{eqnarray}\label{GBh12}
&&h_1(r)=\frac{r+1}{r} \log (\frac{r+1}{r})-\frac{6r^2+3r-1}{6r^3},\\
&&h_2(r)=\frac{r+1}{2r} \log^2 (\frac{r+1}{r})-\frac{6r^2+3r-1}{6r^3} 
\log (\frac{r+1}{r})\nonumber\\
&&\ \ \ \ \ \ \ \ +\frac{5 \left(216 r^3-85 r+27\right) r^2+24
  \left(r \left(r \left(360 r^3-155
  r+69\right)+4\right)+20\right)\alpha }{2160  r^7 (1+8 \alpha)},\\
&& \cdots ,\nonumber
\end{eqnarray}
where we have obtained solutions up to $h_5(r)$. For simplicity we do
not list them here.

From eqs.(\ref{GBh},\ref{GBhsolution},\ref{GBh12}), we can derive
$k(r)$. Expanding $k(r)$ in large $r$, we find
\begin{eqnarray}\label{GBklarger}
k(r)=1-\frac{1}{2r^2}+\frac{\beta_n}{r^4}+O(\frac{1}{r^6})
\end{eqnarray}
where 
\begin{eqnarray}\label{GBbeta}
&&\beta_n=-\frac{1}{8}+\frac{n-1}{12}-\frac{(67+600 \alpha ) }{432
    (1+8 \alpha)} (n-1)^2 
+\frac{\left(151104 \alpha^2+34320 \alpha +1945\right)}
{7776 (1+8 \alpha)^2}  (n-1)^3 \nonumber\\
&& -\frac{\left(1362415104 \alpha^3+471579456 \alpha^2+54244296
  \alpha +2074355\right) }{5598720 (8 \alpha +1)^3} (n-1)^4 \nonumber\\
&& +\frac{\left(19865723572224 \alpha ^4+9304662564864 \alpha^3
+1627900276608 \alpha^2+126143146752 \alpha
+3654194425\right) }{7054387200 (1+8 \alpha )^4}  (n-1)^5\nonumber\\
&&+O(n-1)^6 .
\end{eqnarray}

Recall that $\delta \log Z \sim T^{ij}\delta \gamma_{ij}$ 
and \eq{partitionfirstorder} 
is calculated  on the boundary with $r\to \infty$. 
Thus, $k(r)$ in the large $r$ expansion is good enough for our
purpose. Substituting
eqs.(\ref{GBklarger},\ref{GBmetricK},\ref{GBf},\ref{GBrh}) into
eqs.(\ref{GBTij},\ref{partitionfirstorder},\ref{Renyifirstorderfb}),
we 
obtain
\begin{eqnarray}\label{GBfb1}
&&\frac{8G_N}{\pi}f_b(n)=(1+8\alpha)-\left( \frac{11}{12}+10
  \alpha\right) (n-1)
+\frac{\left(27840 \alpha ^2+5552 \alpha +275\right)
}{216(1+8\alpha)}(n-1)^2
\nonumber\\
&&
-\frac{\left(237436416 \alpha ^3+74097984 \alpha ^2+7667464 \alpha
  +263115\right) }{155520 (1+8 \alpha )^2}(n-1)^3\nonumber\\
&&+\frac{\left(3323533971456 \alpha ^4+1425617289216 \alpha
  ^3+228089069952 \alpha ^2+16137500288 \alpha +426115725\right)
  }{195955200 (1+8 \alpha)^3} (n-1)^4\nonumber\\
&& +O(n-1)^5 .
\end{eqnarray}
Similar to $f_c(n)$, we can rewrite $f_b(n)$ in terms of $a$ and $c$
or $c$ and $t_2$. We have
\begin{eqnarray}\label{GBfb}
f_b(n)&=&c+(\frac{1}{3}a-\frac{5}{4}c)(n-1)+O(n-1)^2\nonumber\\
&=&c+c(-\frac{11}{12}-\frac{1}{18}t_2)(n-1)+O(n-1)^2.
\end{eqnarray}

To end this section, we notice an interesting property of solutions to
GB gravity (\ref{GBaction}). Expanding in $(n-1)$, we find the
solutions such as $f(r)$ and $h(r)$ are exactly the same as those of
Einstein gravity at the first order $(n-1)$. Differences appear only
at higher orders. As we will prove in the next section, this is a
universal property for general higher curvature gravity as long as we
rescale the coefficient of $R$ as $1$.

\subsection{General Higher Curvature Gravity}

In this section, by applying the methods illustrated in sect.2.1,  we
discuss the universal terms of R\'enyi entropy for CFTs dual to
general higher curvature gravity. In general, it is difficult to find
the exact black hole solutions in higher derivative gravity. Instead,
we focus on perturbative solutions up to $(n-1)^2$. This is sufficient
to derive $f_a$ of order $(n-1)^2$ and $f_b, f_c$ of order $(n-1)$. As
we have argued above, it is expected that there is no universal
behavior at higher orders, due to the fact that the higher orders are
determined by four and higher point functions of stress tensor, which
depend on the details of CFTs.

Let us consider the general higher curvature gravity
$I(R_{\mu\nu\rho\sigma})$. We use the trick introduced in
\cite{Miao:2013nfa} to rewrite it into the form similar as
eq.(\ref{GBaction}).  This method together with \cite{Imbimbo:1999bj,Schwimmer:2008yh} is found to be useful to study the
holographic Weyl anomaly and universal terms of entanglement entropy
\cite{Miao:2013nfa,
Miao:2015iba,Miao:2015dua,Sen:2014nfa}
 \footnote{For recent discussions on entanglement entropy and the scale  invariance, please see \cite{Naseh:2016maw}.}. Firstly, we define a
'background-curvature' (we set the AdS radius $l=1$ below)
\begin{eqnarray}\label{backcurvature}
\tilde{R}_{\mu\nu\sigma\rho}=g_{\mu\rho}g_{\nu\sigma}-g_{\mu\sigma}g_{\nu\rho}
\end{eqnarray}
and denote the difference between the curvature and the
'background-curvature' by
\begin{eqnarray}\label{diffcurvature}
\bar{R}_{\mu\nu\sigma\rho}=R_{\mu\nu\sigma\rho}-\tilde{R}_{\mu\nu\sigma\rho}.
\end{eqnarray}
Then we expand the action around this 'background-curvature' and get
\begin{eqnarray}\label{GHCaction0}
I&=&\frac{1}{16\pi G_N}\int d^{d+1}x\sqrt{g} L(R_{\mu\nu\sigma\rho})\nonumber\\
&=&\frac{1}{16\pi G_N}\int d^{d+1}x\sqrt{g}\big[ L_0+ c^{(1)}_1
  \bar{R} +( c^{(2)}_1 \mathcal{L}_4(\bar{R})+ c^{(2)}_2
  \bar{R}_{\mu\nu}\bar{R}^{\mu\nu}+ c^{(2)}_3\bar{R}^2)+\sum_{\ii=1}^8
  c^{(3)}_\ii \mathcal{K}_\ii(\bar{R})+O(\bar{R}^4) \big]\nonumber\\
\end{eqnarray}
where
$L_0=L(\tilde{R}_{\mu\nu\sigma\rho})=L(R_{\mu\nu\sigma\rho})|_{AdS}$ is a constant defined by
 the Lagrangian for AdS solution, and
 $c^{(n)}_\ii$ are constants which parametrize the higher derivatives
correction to the Einstein action up to third orders in the curvature
with $n$ denoting the order.
Here 
$\mathcal{L}_4(\bar{R})$ denotes the GB term
\begin{eqnarray}\label{BGterm}
\mathcal{L}_4(\bar{R})=
\bar{R}_{\mu\nu\rho\sigma}\bar{R}^{\mu\nu\rho\sigma}-
4\bar{R}_{\mu\nu}\bar{R}^{\mu\nu}+\bar{R}^2, 
\end{eqnarray}
and 
$\mathcal{K}_\ii(\bar{R})$ denotes the basis of third order curvature terms
\begin{eqnarray}\label{Kni}
&&\mathcal{K}_\ii(\bar{R})=
\{ 
\bar{R}^3,\bar{R}\bar{R}_{\mu\nu}\bar{R}^{\mu\nu},\bar{R}\bar{R}_{\mu\nu\rho\sigma}
\bar{R}^{\mu\nu\rho\sigma},\bar{R}_{\mu}^{\nu}\bar{R}_{\nu}^{\rho}\bar{R}_{\rho}^{\mu},
\bar{R}^{\mu\nu}\bar{R}^{\rho \sigma}\bar{R}_{\mu \rho \sigma\nu},
\bar{R}_{\mu \nu}\bar{R}^{\mu \rho \sigma
  \lambda}\bar{R}^{\nu}_{\ \rho \sigma \lambda},\nonumber\\
&&\ \ \ \ \ \ \ \ \ \ \ \ \ \bar{R}_{\mu \nu \rho \sigma}\bar{R}^{\mu \nu
\lambda\chi}\bar{R}^{\rho \sigma}_{\ \ \lambda\chi},\bar{R}_{\mu \nu
\rho\sigma}\bar{R}^{\mu\lambda\chi\sigma}\bar{R}^{\nu\ \ \rho}_{\
\lambda\chi} \}.
\end{eqnarray}
We require that the higher derivative gravity has an asymptotic AdS solution. This would impose a condition $ c^{(1)}_1=-L_0/2d$ \cite{Miao:2013nfa}. Using this condition, we can rewrite the action (\ref{GHCaction0}) as
\begin{eqnarray}\label{GHCaction1}
I=\frac{1}{16\pi G_N}\int_M 
-\frac{L_0}{2d}(R+d^2-d) +( c^{(2)}_1 \mathcal{L}_4(\bar{R})+
  c^{(2)}_2 \bar{R}_{\mu\nu}\bar{R}^{\mu\nu}+
  c^{(2)}_3\bar{R}^2)+\sum_{\ii=1}^8 c^{(3)}_\ii
  \mathcal{K}_\ii(\bar{R})+O(\bar{R}^4)  \ .
\end{eqnarray}
Rescaling $G_N\to \tilde{G}_N= -\frac{2d}{L_0}G_N$, $c^{(n)}_i\to \tilde{c}^{(n)}_i=-\frac{2d}{L_0}c^{(n)}_i $, we have
\begin{eqnarray}\label{GHCaction2}
I=\frac{1}{16\pi\tilde{ G}_N}\int_M 
(R+d^2-d) +( \tilde{c}^{(2)}_1 \mathcal{L}_4(\bar{R})+
\tilde{c}^{(2)}_2 \bar{R}_{\mu\nu}\bar{R}^{\mu\nu}+
\tilde{c}^{(2)}_3\bar{R}^2)+\sum_{\ii=1}^8 \tilde{c}^{(3)}_\ii
\mathcal{K}_\ii(\bar{R})+O(\bar{R}^4) .
\end{eqnarray}
Now it takes the form as eq.(\ref{GBaction}).  For simplicity, 
we ignore the notation $ \tilde{ }\;$ below. The 
E.O.M of the above gravity is
\begin{eqnarray}\label{HDEOM}
P_{\mu}^{\ \alpha\rho\sigma}R_{\nu\alpha\rho\sigma}-2\triangledown^{\rho}
\triangledown^{\sigma}P_{\mu\rho\sigma\nu}-\frac{1}{2}L g_{\mu\nu}=0,
\end{eqnarray}
with $P^{\mu\alpha\rho\sigma}=\partial L/\partial R_{\mu \alpha\rho\sigma}$.
 
A couple of remarks on action (\ref{GHCaction2}) are in order.

Firstly, it is clear the hyperbolic black hole which is locally AdS is
a solution to action (\ref{GHCaction2}). That is because
$\bar{R}_{\mu\nu\rho\sigma}=0$ in AdS. We are interested of two kinds
of perturbations: the first one is $\delta g_{\mu\nu}\sim O(n-1)$
related to $f_a, f_c$, and the second one is $\delta g_{\mu\nu}\sim
O\big{(}(n-1),K\big{)}$ related to $f_b$. Remarkably, we have
$\bar{R}_{\mu\nu\rho\sigma}\sim O(n-1, K^2)$ \footnote{ Note that we
  have  $\bar{R}_{\mu\nu\rho\sigma}$ proportional to $O( K^2)$ instead
  of $O( K)$. The reason is as follows: $K$ depends on the
  orientation, while $R$ is orientation independent. Thus $R$ must be
  proportional to even powers of $K$.  Substituting $f(r)=r^2-1$ and
  $k(r)=\sqrt{r^2-1}/r$ into the metric (\ref{GBmetricK}), one can
  check that indeed $\bar{R}_{\mu\nu\rho\sigma}\sim O(K^2)$.  } for
the deformed metric (\ref{GBmetricK}) .

Secondly, we are interested of the solutions up to $O(n-1)^2$ and
$O(K)$, or equivalently, the action up to $O(n-1)^3$ and
$O\big{(}(n-1)^2 K^2\big{)}$. As a result, we can drop the
$O(\bar{R})^4$ terms in action (\ref{GHCaction2}) due to
$O(\bar{R})^4\sim O\big{(}(n-1)^4,(n-1)^3 K^2, \cdots \big{)}$. 
Recall that the terms of order $O\big{(} (n-1)^a K^b\big{)}$ in the
action contributes to terms at least of order  $O\big{(} (n-1)^{a-1}
K^b\big{)}$ and  $O\big{(} (n-1)^{a} K^{b-1}\big{)}$ in
the E.O.M.

Thirdly, at the linear order in $O(n-1, K)$, solutions to Einstein
gravity are also solutions to higher curvature gravity
(\ref{GHCaction2}). In other words, the parameters $\tilde{c}^{(n)}_i$
do not appear in the solutions of order $O(n-1,K)$. Let us give a
simple proof.
Since $\mathcal{K}_i(\bar{R})\sim\bar{R}^3\sim
O(n-1)^3\ \text{and}\ O\big{(}(n-1)^2 K^2\big{)}$, 
obviously 
they do
not contribute to the solution at order $O(n-1,K)$. Now we are left
with three curvature-squared terms. Notice that $ \bar{R}_{\mu\nu}=0$
and $ \bar{R}=0$ for all solutions to Einstein gravity with negative
cosmological constant. Thus we only need to consider the GB term
$\mathcal{L}_4(\bar{R})$, which contributes the following terms to
the E.O.M
\begin{eqnarray}\label{GHEOM}
P_{\mu}^{\ \alpha\rho\sigma}R_{\nu\alpha\rho\sigma}-\frac{1}{2}\mathcal{L}_4(\bar{R})
g_{\mu\nu},
\end{eqnarray}
where $P^{\mu\alpha\rho\sigma}=\partial\mathcal{L}_4(\bar{R})/\partial
R_{\mu \alpha\rho\sigma}$.
At leading order we have
$P_{\mu}^{\ \alpha\rho\sigma}R_{\nu\alpha\rho\sigma}\sim 2
\bar{R}_{\mu}^{\ \alpha\rho\sigma}R_{\nu\alpha\rho\sigma}\sim
4\bar{R}_{\mu\nu}\sim O\big{(}(n-1)^2, K^2\big{)}$ and
$\mathcal{L}_4(\bar{R})\sim O\big{(}(n-1)^2, K^2\big{)}$. Thus, it is
clear that  the GB term $\mathcal{L}_4(\bar{R})$ does not affect the E.O.M
of order $O(n-1,K)$. This is indeed the case as we have seen in
sec. 2.1.  Now we finish the proof.

Finally, let us discuss the regularized boundary stress tensor of
action (\ref{GHCaction2}). Let us firstly discuss the
curvature-squared terms in action (\ref{GHCaction2}). Such terms are
studied in \cite{Cremonini:2009ih} at the first order of $c^{(2)}_2$
and $c^{(2)}_3$.  Reparameterizing their formulas, we find for
the Brown-York boundary stress tensor
\begin{eqnarray}\label{GHCTij}
8\pi \tilde{G}_NT_{\partial M}^{ij}&=&(1+2{c}^{(2)}_1(d-1)(d-2)) \big{[}
  K_{\partial M}^{ij}-K_{\partial M} \gamma^{ij}-(d-1)\gamma^{ij}+\frac{\Theta(d-3)}{d-2}
  (R_{\partial M}^{ij}-\frac{1}{2}R_{\partial M} \gamma^{ij})
  \big{]}\nonumber\\
&&+2{c}^{(2)}_1 (Q^{ij}-\frac{1}{3}Q \gamma^{ij}),
\end{eqnarray}
where $d=4$ and $Q_{ij}$ is given by eq.(\ref{GBQij}). 
Remarkably, 
the terms 
$c^{(2)}_2 \bar{R}_{\mu\nu}\bar{R}^{\mu\nu}$ and $c^{(2)}_3\bar{R}^2$
do not contribute to the  regularized boundary stress
tensor. 
This is actually expected since for an asymptotically
AdS spacetime, 
we can rewrite the metric in Fefferman-Graham gauge
\begin{eqnarray}\label{FG}
ds^2=g_{\mu\nu}dx^{\mu}dx^{\nu}=\frac{1}{4\hat{\rho}^2}d\hat{\rho}^2+\frac{1}{\hat{\rho}}\gamma_{ij}dx^idx^j,
\end{eqnarray}
where  $\gamma_{ij}=\gamma_{(0)ij}+\hat{\rho} \gamma_{(1)ij}+ \cdots$
and the boundary is at $\hat{\rho}\to 0$.  Near the boundary, we have
\cite{Miao:2013nfa}
\begin{eqnarray}\label{asymptoticalbehaves1}
&&\sqrt{g} R \sim \sqrt{g} \sim O(\frac{1}{\hat{\rho^3}}),\\ \label{asymptoticalbehaves2}
&&\sqrt{g}  \mathcal{L}_4(\bar{R}) \sim O(\frac{1}{\hat{\rho}}),\\   \label{asymptoticalbehaves3}
&&\sqrt{g} \mathcal{K}_7(\bar{R})\sim \sqrt{g} \mathcal{K}_8(\bar{R}) \sim O(1),\\ \label{asymptoticalbehaves4}
&&\sqrt{g}  \bar{R}_{\mu\nu}\bar{R}^{\mu\nu} \sim \ \sqrt{g}  
\bar{R}\bar{R}\sim \sqrt{g} \mathcal{K}_{\ii \ne
  7,8}(\bar{R})\sim\sqrt{g}  
O(\bar{R}^4)  
\sim O(\hat{\rho}).
\end{eqnarray}
Clearly, only terms
(\ref{asymptoticalbehaves1},\ref{asymptoticalbehaves2}) in action
(\ref{GHCaction2}) are divergent and need to be regularized near the
boundary. No counter terms are needed for the other terms 
for 
$d=4$. In addition to the counter terms which make the action finite,
one may worry about the Gibbons-Hawking-York 
(GHY) boundary terms which make
a well-defined variational principle. For general higher curvature
gravity, the GHY-like term is proposed in \cite{Deruelle:2009zk}.
For $ \mathcal{K}_{i}(\bar{R})$, we have
\begin{eqnarray}\label{GHYterm}
I_{\text{GHY}}\sim \int_{\partial M} 
d^4x
\frac{
  \sqrt{\gamma}}{\hat{\rho}^2}P^{\hat{\rho} i}_{\ \ \hat{\rho} j}
K_{\partial M \ i}^{\ \ \ \ j}\sim O(\hat{\rho}),
\end{eqnarray}
where $P^{\mu\nu\rho\sigma}=\partial \mathcal{K}_{\ii}(\bar{R}) /
\partial R_{\mu\nu\rho\sigma}$. So the GHY-like terms for $
\mathcal{K}_{i}(\bar{R})$ are harmless. The GHY-like terms and counter
terms for curvature-squared are discussed in \cite{Cremonini:2009ih},
which yield eq.(\ref{GHCTij}).

 In conclusion, the regularized boundary stress tensor for higher
 curvature gravity (\ref{GHCaction2}) is given by eq.(\ref{GHCTij}) in
 dimensions less than five ($d=4$).   It should be stressed that the
 GHY-like terms and counter terms for $\mathcal{K}_7(\bar{R})$ and $
 \mathcal{K}_8(\bar{R})$ are necessary when $d\ge 6$.

\subsubsection{$f_a(n)$}

Applying the methods of sect. 2.1.1, let us calculate $f_a(n)$ in
general higher curvature gravity (\ref{GHCaction2}). Recall that
R\'enyi entropy on spherical entangling surface is given by
\begin{eqnarray}\label{GHCRenyi}
S_n=\frac{n}{1-n}\frac{1}{T_0}\int_{T_0/n}^{T_0}S_{\text{BH}}(T)dT
\end{eqnarray}
with $S_{\text{BH}}(T)$ the black hole entropy
\begin{eqnarray}\label{GHCBHentropy}
S_{\text{BH}}=\frac{1}{8G_N}\int_H dy^3\sqrt{h}\frac{\partial L}
{\partial R_{\mu\nu\rho\sigma}}\varepsilon_{\mu\nu}\varepsilon_{\rho\sigma}.
\end{eqnarray}

To suppress the massive modes and ghost modes with $M\sim 1/ c^{(n)}_\ii
$, we work in perturbative framework with $c^{(n)}_\ii \ll 1$. After
some calculations, we find the black hole solution as
\begin{eqnarray}
\label{GBmetricF}
ds^2_{\text{bulk}}=\frac{dr^2}{f(r)}+f(r)F(r)d\tau^2+r^2 d\Sigma^2_3
\end{eqnarray}
where $ d\Sigma^2_3$ is the line element for hyperbolic plane $H^3$ with unit curvature, and $f(r), F(r)$ are given by
\begin{eqnarray}\label{GHCf}
f(r)&=&r^2-1+\frac{2 (n-1)}{3 r^2}\nonumber\\
&&-\frac{ \left(r^6 (336 c^{(2)}_1+192 c^{(3)}_7-96 c^{(3)}_8+35)-24
  r^2 
( c^{(2)}_1+228 c^{(3)}_7-3 c^{(3)}_8)+4608 c^{(3)}_7\right)}{27 (1+8
  c^{(2)}_1) r^8}(n-1)^2, \nonumber\\
&&+O(n-1)^3\\\label{GHCF}
F(r)&=&1-\frac{8 (52c^{(3)}_7+3 c^{(3)}_8)}{3 (1+8 c^{(2)}_1) r^8}(n-1)^2 +O(n-1)^3.
\end{eqnarray}
From the conditions 
\begin{eqnarray}\label{GHCHawkingT}
&& f(r_H)=0,\\
&&T=\frac{1}{4\pi}\sqrt{f'(r)\partial_r [f(r)F(r)]}|_{r=r_H}=\frac{1}{2\pi n},
\end{eqnarray}
we find a consistent solution
\begin{eqnarray}\label{GHCrh}
r_H=1-\frac{(n-1)}{3}+\frac{10}{27} (n-1)^2+\frac{4  (4
  c^{(2)}_1-84c^{(3)}_7-3c^{(3)}_8)}
{27 (1+8c^{(2)}_1)}(n-1)^2+O(n-1)^3.
\end{eqnarray}
Substituting the above equations into eqs.(\ref{GHCRenyi},
\ref{GHCBHentropy}), we obtain
\begin{eqnarray}\label{GHCfa1}
\frac{8G_N}{\pi}f_a(n)=1-\frac{1+8c^{(2)}_1}{2} (n-1)+\frac{1}{54}
(336 c^{(2)}_1+192 c^{(3)}_7-96 c^{(3)}_8+35)(n-1)^2+ \cdots .
\end{eqnarray}

Using the following relations \cite{Sen:2014nfa}, 
\begin{eqnarray}\label{Indiaac}
a&=&\frac{\pi}{8G_N},\quad  c=\frac{\pi}{8G_N}(1+8c^{(2)}_1),\\ \label{Indiat2}
t_2&=&\frac{12}{1+8c^{(2)}_1}(4c^{(2)}_1-192c^{(3)}_7+96c^{(3)}_8),\\ \label{Indiat4}
t_4&=&\frac{2160}{1+8c^{(2)}_1}(2c^{(3)}_7-c^{(3)}_8),
\end{eqnarray}
we 
can rewrite eq.(\ref{GHCfa1}) as
\begin{eqnarray}\label{GHCfa2}
&&f_a(n)=a -\frac{c}{2}(n-1)+ c(\frac{35}{54}+\frac{7}{324}t_2+\frac{1}{84}t_4)(n-1)^2+\cdots
\end{eqnarray}
which is eq.(\ref{mainresultfa}) advertised in the Introduction.

We remark that although we work in linear order of $c^{(n)}_\ii $ in the
above derivation, our result eq.(\ref{GHCfa2}) applies to finite
$c^{(n)}_\ii $. For the case $c^{(2)}_2=c^{(2)}_3=0 $,
eqs.(\ref{GHCf},\ref{GHCF},\ref{GHCrh}) are exact in $c^{(n)}_\ii $. For
small but non-zero $c^{(2)}_2$ and $c^{(2)}_3$, we have performed a
fifth order perturbation and find that eq.(\ref{GHCfa2}) remains
unchanged.

\subsubsection{$f_c(n)$}

Now let us study $f_c(n)$ in higher curvature gravity
(\ref{GHCaction2}). 
Similarly, we consider the first order variation
\eq{partitionfirstorder}
of the partition 
function with $T_{ij}$ computed on the undeformed  hyperboloid
background.
Here $T^{ij}$ is the regularized boundary stress tensor given by
eq.(\ref{GHCTij}). The bulk metric takes the form
\begin{eqnarray}\label{GBmetricQF}
ds^2_{\text{bulk}}=\frac{dr^2}{f(r)}+f(r)F(r)d\tau^2+\frac{r^2}{\rho^2}
\big{[}d\rho^2+(\delta_{\hat{i}\hat{j}}+q(r)Q_{ab\hat{i}\hat{j}}x^ax^b+O(\rho^3))dy^\ih
  dy^\jh \big{]},
\end{eqnarray}
which approaches the deformed hyperboloid eq.(\ref{deformedmetricQ})
for $q(\infty)=1$. 
Recalling  $f(r), F(r)$ as given in
eqs.(\ref{GHCf},\ref{GHCF}) and substituting all these equations together
with $\gamma_{\hat{i}\hat{j}}=\frac{r^2}{\rho^2}Q_{ab\hat{i}\hat{j}}x^ax^b$
into eqs. (\ref{partitionfirstorder},\ref{Renyifirstorder}),
we obtain
\begin{eqnarray}\label{GHCfc1}
\frac{8G_N}{\pi}f_c(n)=(1+8c^{(2)}_1) +  (-\frac{17}{18}-\frac{32}{3}
c^{(2)}_1-\frac{32}{3} c^{(3)}_7+\frac{16}{3} c^{(3)}_8)(n-1)+\cdots.
\end{eqnarray}
Applying eqs.(\ref{Indiaac},\ref{Indiat2},\ref{Indiat4}), we 
can rewrite $f_c(n)$ as
\begin{eqnarray}\label{GHCfc2}
f_c(n)=c+ c
(-\frac{17}{18}-\frac{7}{108}t_2-\frac{1}{27}t_4)(n-1)+\cdots . 
\end{eqnarray}

Notice that $f_a(n)$ (\ref{GHCfa2}) and $f_c(n)$ (\ref{GHCfc2}) are
consistent with identity (\ref{fafc}). This is a non-trivial check of
our holographic approach, in particular, the regularized boundary
stress tensor eq.(\ref{GHCTij}).

\subsubsection{$f_b(n)$}

Finally, let us discuss $f_b(n)$ in the higher derivative gravity. 
Similar to the case of the GB gravity, 
the key point is to find deformed black hole solutions 
up to order  $O(K)$
\begin{eqnarray}\label{GHCmetricK}
ds^2_{\text{bulk}}=\frac{dr^2}{f(r)}+f(r)F(r)d\tau^2+\frac{r^2}{\rho^2}
\big{[}d\rho^2+(\delta_{\hat{i}\hat{j}}+k(r)K_{a\hat{i}\hat{j}}x^a+O(\rho^2))dy^i
  dy^j \big{]}
\end{eqnarray}
 For traceless $K_{aij}$, there is one independent equation of
 $k(r)$. We find the solution at the linear order in $(n-1)$ is
 exactly the same as that of Einstein gravity, which agrees with the
 arguments below eq.(\ref{HDEOM}). Modifications from the
 higher-curvature terms only appear at higher orders. Remarkably, at
 the next order $O(n-1)^2$, only $c^{(2)}_1, c^{(3)}_7$ and $
 c^{(3)}_8$ contribute.

Following the approach of sect. 2.1.3, we obtain $k(r)$ in large $r$ expansion
as
\begin{eqnarray}\label{GHCklarger}
k(r)=1-\frac{1}{2r^2}+\frac{\beta_n}{r^4}+O(\frac{1}{r^6}),
\end{eqnarray}
where 
\begin{eqnarray}\label{GHCbeta}
\beta_n=-\frac{1}{8}+\frac{n-1}{12}+\frac{ (-600 c^{(2)}_1+4224
  c^{(3)}_7-2112  c^{(3)}_8-67)}{432 (1+8c^{(2)}_1)}(n-1)^2+O(n-1)^3.
\end{eqnarray}
Substituting eqs.(\ref{GHCklarger},\ref{GHCf},\ref{GHCF}) and 
$\delta
\gamma_{\hat{i}\hat{j}}=\frac{r^2}{\rho^2}K_{a\hat{i}\hat{j}}x^a$ into
\eq{partitionfirstorder} and \eq{Renyifirstorderfb}, we
obtain
\begin{eqnarray}\label{GHCfb1}
f_b(n)&=&\frac{\pi}{8G_N}\big{[}(1+8c^{(2)}_1) +  (-\frac{11}{12}-10 
c^{(2)}_1+32c^{(3)}_7-16 c^{(3)}_8)(n-1)+\cdots\big{]}\\ \label{GHCfb2}
&=& c-c( \frac{11}{12}+\frac{1}{18}t_2+\frac{1}{45}t_4 )(n-1)+\cdots
\end{eqnarray}
as declared in the Introduction.

Now we have obtained $f''_a(1), f'_b(1)$ and $f'_c(1)$ by using
holographic methods. Interestingly, they only depend on the parameters
of stress tensor two-point and three-point functions. When
$c^{(2)}_2=c^{(2)}_3=0$, our derivations are nonperturbative in the
coupling constants of higher curvature gravity. For small but non-zero
$c^{(2)}_2$ and $c^{(2)}_3$, we have performed a fifth order
perturbation and find that they remains unchanged. In conclusion,
our obtained results
eqs.(\ref{mainresultfa},\ref{mainresultfb},\ref{mainresultfc}) are
universal laws for strongly coupled CFTs 
that are dual to general higher
curvature gravity.
It is expected that there are no such universal laws at the next
order, since 
the next order terms would involve the stress energy 
four-point functions which no longer admit any universal form.

\section{The Story of Free CFTs}

In this section, we discuss the universal terms of R\'enyi entropy for
free CFTs. We find the holographic relations found in sect.2 also
apply to free fermions and free vectors but not to free scalars. We
find a combined relation which is obeyed by all free CFTs and strongly
coupled CFTs with holographic dual. It seems that this  combined
relation is universal for all CFTs in four dimensions.

For the theory consisting of $n_s$ free real scalars, $n_f$ free Weyl fermions and $n_v$ free vectors, the functions $f_a(n)$ and $f_c(n)$ have been calculated explicitly in \cite{Bueno:2015qya,Fursaev:2012mp,Klebanov:2011uf,Casini:2010kt,Fursaev:1993hm,DeNardo:1996kp}. We list the results as follows:
\begin{eqnarray}\label{Freefa}
f_a(n)&=&\frac{1}{360}\big{[}  n_s \frac{(1+n)(1+n^2)}{4n^3}+n_f \frac{(1+n)(7+37n^2)}{16n^3}+n_v \frac{(1+n+31n^2+91n^3)}{2n^3} \big{]},\\ \label{Freefc}
f_c(n)&=&\frac{1}{120}\big{[}  n_s \frac{(1+n)(1+n^2)}{4n^3}+n_f \frac{(1+n)(7+17n^2)}{16n^3}+n_v \frac{(1+n+11n^2+11n^3)}{2n^3} \big{]}.
\end{eqnarray}
One can check that the above $f_a(n)$ and $f_c(n)$ satisfy the identity eq.(\ref{fafc}). Assuming $f_b(n)=f_c(n)$, we have
\begin{eqnarray}\label{Freefb}
f_b(n)&=&\frac{1}{120}\big{[}  n_s \frac{(1+n)(1+n^2)}{4n^3}+n_f \frac{(1+n)(7+17n^2)}{16n^3}+n_v \frac{(1+n+11n^2+11n^3)}{2n^3} \big{]}.
\end{eqnarray}
This is at least the case for free scalars \cite{Bianchi:2015liz}.  Numerical calculations also support $f_b(n)=f_c(n)$ for free fermions \cite{Lee:2014xwa}. 

According to \cite{Osborn:1993cr,Erdmenger:1996yc}, the stress tensor
three-point functions for CFTs 
in general spacetime dimensions
are completely determined in terms of the three
parameters $A, B, C$ as, 
\begin{eqnarray}
\label{CT0}
C_T&=&\frac{\pi^{d/2}}{d(d+2)\Gamma[d/2]}[ (d-1)(d+2)A-2B-4(d+1)C ], 
\\ \label{t2}
t_2&=&\frac{2(d+1)}{d} \frac{(d-2)(d+2)(d+1)A+3d^2
  B-4d(2d+1)C}{(d-1)(d+2)A-
2B-4(d+1)C}, \\ \label{t4}
t_4&=&-\frac{(d+1)}{d} \frac{(d+2)(2d^2-3d-3)A+2d^2(d+2) B-
4d(d+1)(d+2)C}{(d-1)(d+2)A-2B-4(d+1)C},
\end{eqnarray}
where for  free 4d CFTs, we have
\begin{eqnarray}\label{FreeA}
A&=&\frac{8}{27\pi^6} (n_s-54 n_v), \\ \label{FreeB}
B&=&-\frac{2}{27\pi^6} (8n_s+432n_v+27 n_f), \\ \label{FreeC}
C&=&-\frac{1}{27\pi^6} (2n_s+432 n_v+27 n_f). 
\end{eqnarray}
and $C_T=\frac{40}{\pi^4}c$.

Substituting eqs.(\ref{FreeA}-\ref{t4}) into the holographic relations
eqs.(\ref{mainresultfa},\ref{mainresultfb},\ref{mainresultfc})  for
$f_a, f_b, f_c$  and comparing with those of free CFTs
eqs.(\ref{Freefa},\ref{Freefc},\ref{Freefb}), we find exact agreements
for fermions and vectors. However, there is discrepancy for
scalars. As noticed in \cite{Lee:2014zaa}, such discrepancy results
from the boundary contributions to the modular Hamiltonian.
Interestingly, we find the following combined holographic relations
\begin{eqnarray}\label{Conjecturefbfc}
&&2f'_b(1)-3f'_c(1)=c(1+\frac{1}{12}t_2+\frac{1}{15}t_4)\\ \label{Conjecturefbfa}
&&2f'_b(1)+\frac{9}{2}f''_a(1)=c(4+\frac{1}{12}t_2+\frac{1}{15}t_4)
\end{eqnarray}
are satisfied by all free CFTs including scalars. We conjecture 
these are universal laws for all CFTs in four dimensions. As mentioned
in the
Introduction, eq.(\ref{Conjecturefbfc}) and eq.(\ref{Conjecturefbfa}) 
are not independent, which can be derived from each other by applying 
eq.(\ref{fafc}).  

In the notation of \cite{Bianchi:2015liz}, our conjecture
(\ref{Conjecturefbfc}) becomes
\be
\pi  C_D''(1)-36 h_n''(1)
=\frac{2\pi^3}{5} C_T(1+\frac{1}{12}t_2+\frac{1}{15}t_4), 
\quad \mbox{(for 4d CFTs),}
\ee
where $C_T=\frac{40}{\pi^4}c$ for 4d.
As the quantities $h_n$ and $C_D$ have natural
definitions in all dimensions. It is expected that one can generalize
our results to general dimensions. We will perform this analysis in
the next section.

\section{Universality of HRE in General Dimensions}

In this section, we study $h_n(n)$ and $C_D(n)$ of holographic R\'enyi
entropy 
for CFT in general $d$-dimensions. We firstly 
consider the 3d case 
and then discuss the case in higher dimensions. We find that 
in general dimensions 
there are indeed 
similar holographic universal laws expressing 
$h_n''(1)$ and $C_D''(1)$  in terms of  a linear combination of $C_T, t_2$
and $t_4$. And for all the examples we have checked, these holographic
laws are obeyed by free fermions, but are violated by free
scalars. Similar to what we did above for four dimensions, we are also
able to find a specific relation involving linearly the quantities 
$h_n''(1), C_D''(1), C_T, t_2$ and $t_4$, 
which applies to free fermions, free scalars and
strongly coupled CFTs with holographic dual. We conjecture 
that this
relation holds for general CFTs.

To proceed,
we apply the holographic approach developed in
\cite{Dong123,Balakrishnan:2016ttg} to derive $h_n(n)$ and $C_D(n)$
for general higher curvature gravity. 
This procedure treats the extrinsic curvature perturbatively. For our purpose, we only need to consider the linear order of the extrinsic curvature below.
 Inspired by
\cite{Dong123,Balakrishnan:2016ttg}, we consider the following bulk
metric 
\begin{eqnarray}\label{generalbulkmetric}
ds^2_{\text{bulk}}&=&\frac{dr^2}{f(r)}+f(r)F(r)d\tau^2\nonumber\\
&+&\frac{r^2}{\rho^2}
\big[ d\rho^2+(\delta_{\hat{i}\hat{j}}+2k(r)\bar{K}_{a\hat{i}\hat{j}}x^a)dy^\ih
  dy^\jh +\frac{4}{d-2}k(r)\partial_i K^ax_a \rho d\rho dy^\ih  +O(\rho^2)\big],
\end{eqnarray}
where $\bar{K}_{a\hat{i}\hat{j}}$ is the traceless part of 
extrinsic curvature and we have $\partial_{\hat{k}}
K_{a\hat{i}\hat{j}}=\partial_{\hat{j}} K_{a\hat{i}\hat{k}}+O(K^2)$ for
consistency \cite{Dong123}. According to
\cite{Dong123,Balakrishnan:2016ttg,Billo:2016cpy}, $h_n$ and $C_D(n)$ can be
extracted from the boundary stress tensor
\begin{eqnarray}\label{hnCDfromTij}
&& \la T_{ab}(x) \ra_n=\frac{g_n}{\rho^2}\big{(}   (d-1) \delta_{ab}-d 
\frac{x_ax_b}{\rho^2} \big{)}+\cdots,\nonumber\\
&& \la T_{a\hat{i}}(x)
\ra_n=\frac{x_ax_b}{\rho^2}\partial_{\hat{i}}K^b \frac{k_n}{d-2}
+\cdots,\\
&&\la T_{\hat{i}\hat{j}}(x) \ra_n=\frac{1}{\rho^2}\big{(}  -g_n
\delta_{\hat{i}\hat{j}}+k_n \bar{K}^a_{\hat{i}\hat{j}}
x_a\big{)}+\cdots,\nonumber
\end{eqnarray}
where 
\begin{eqnarray}\label{kngn}
k_n-k_1=\frac{(d-1)\Gamma(\frac{d}{2}-1)\pi^{\frac{d}{2}-2}}
{2\Gamma(d+1)}\frac{C_D}{n}-\frac{3d-4}{d-2}\frac{h_n}{2\pi  n},
\quad \mbox{and}\quad 
  g_n-g_1=\frac{h_n}{2\pi n}.
\end{eqnarray}
The boundary stress tensor in general higher curvature gravity 
has been 
calculated in \cite{Sen:2014nfa}, yielding that
\begin{eqnarray}\label{hnCDholoTij}
\la T_{ij} \ra =\frac{d}{f_d}C_T h^{(d)}_{ij}=\frac{d}{16\pi
  G_N}\big{(}1+4(d-2)c^{(2)}_1\big{)} h^{(d)}_{ij},
\end{eqnarray}
where 
\be  \label{fd}
f_d=2\frac{d+1}{d-1}\frac{\Gamma(d+1)}{\pi^{d/2}\Gamma(d/2)},
\ee
\be \label{CT}
C_T=\frac{f_d}{16\pi G_N}\big{(}1+4(d-2)c^{(2)}_1\big{)}.
\ee
and $ h^{(d)}_{ij}$ appears in the Fefferman-Graham expansion of the
asymptotic AdS metric
\begin{eqnarray}\label{hnCDFGAdS}
ds^2=\frac{dz^2}{z^2}+\frac{1}{z^2}( g^{(0)}_{ij} + z^2  g^{(1)}_{ij}+\cdots+z^d  h^{(d)}_{ij}+\cdots   )dy^idy^j.
\end{eqnarray}
Notice that the stress-tensor eq.(\ref{hnCDholoTij}) contains
contributions from the $g_{(0)ij}$ in even dimensions
\cite{deHaro:2000vlm}.  
These contributions reflect the presence of conformal
anomalies. However, as argued in \cite{Dong123},  these terms do not
affect $C_D(n)$ and $h_n$ \footnote{One can easily check that
  $g_{(0)ij}$ is independent of $M_e$ and $\beta_n$. Thus, the contributions to
  the stress-tensor eq.(\ref{hnCDholoTij}) from $g_{(0)ij}$ 
do not affect $C_D(n)$ and $\beta_n$. }. So we have ignored them
in the present paper. Note also that we use a seemingly different
stress tensor $ T_{\partial M ij}$  eq.(\ref{GHCTij}) in
sect. 2. Actually,  the stress-tensor  eq.(\ref{GHCTij}) is equivalent
to  eq.(\ref{hnCDholoTij}) up to a rescaling and some functions of
$g_{(0)ij}$ \cite{deHaro:2000vlm}
\begin{eqnarray}\label{relation}
<T_{ij}>=\lim_{z \to 0} \frac{1}{z^{d-2}} T_{\partial M ij}.
\end{eqnarray}
If we take the stress tensor eq.(\ref{hnCDholoTij}) instead of
eq.(\ref{GHCTij}) in the procedure of sect. 2, we get the same results
for $f_b(n)$ and $f_c(n)$. The interested reader is referred to 
Appendix A for the proof of the equivalence. Now let us focus on the
stress tensor eq.(\ref{hnCDholoTij}) from now on.

Comparing eq.(\ref{hnCDholoTij}) with eq.(\ref{hnCDfromTij}), one can
read out $h_n(n)$ and $C_D(n)$. Let us take Einstein gravity as an example. The
solution is given by
\begin{eqnarray}\label{EinfF}
f(r)&=&r^2-1-\frac{M}{r^{d-2}}, \ \ \ F(r)=1 ,\\  \label{Eink}
k(r)&=& \frac{\sqrt{r^2-1}}{r}+\frac{\beta_{n}}{r^d}+O(\frac{1}{r^{d+1}}).
\end{eqnarray}
From the above equations, one can easily obtain 
\begin{eqnarray}\label{Einhij}
 h^{(d)}_{\hat{i}\hat{j}}=\frac{1}{\rho^2}\big{[} (\frac{1}{d}M+ g_0)  \delta_{\hat{i}\hat{j}}+ (\frac{2}{d}M+2\beta_{n}+ k_0) \bar{K}^a_{\hat{i}\hat{j}} x_a   \big{]},
\end{eqnarray}
where $g_0$ and $k_0$ are constants which are not
important \footnote{From
  eqs.(\ref{hnCDfromTij},\ref{hnCDholoTij},\ref{Einhij}), we can
  derive $k_n$ and $g_n$, 
which 
have a linear dependence on the constants  $g_0$ and $k_0$
appearing in eq.(\ref{Einhij}). However, we are interested of $C_D(n)$
and $h_n$ instead of $k_n$ and $g_n$. Since $C_D(n)$ and $h_n$  are
functions of $(k_n-k_1)$ and $(g_n-g_1)$ from eq.(\ref{kngn}). They do
not depend on $g_0$ and $k_0$  instead.}. 
Comparing eqs.(\ref{hnCDholoTij},\ref{Einhij}) with the last equation
of (\ref{hnCDfromTij}), one obtains
\cite{Dong123,Balakrishnan:2016ttg} 
\begin{eqnarray}\label{Einhn}
&&\frac{h_n}{n}=-\frac{M}{8G_N},\\ \label{EinCD}
&& \frac{C_D}{n}=\frac{d\Gamma(d+1)}{(d-1)\pi^{d/2-2}\Gamma(d/2)}\frac{ 2(d-2) (\beta_n-\beta_1)-M}{16\pi G_N}.
\end{eqnarray}

Now let us turn to discuss the general higher curvature gravity
(\ref{GHCaction2}). In general, it is difficult to find the black hole
solutions for higher derivative gravity. For simplicity, we work in
the perturbative framework of the coupling constants $c^{(n)}_i$. 
Remarkably, we find the solutions behaving as
\begin{eqnarray}\label{GHCfF}
f(r)&=&r^2-1-\frac{M_e}{r^{d-2}}+O(\frac{1}{r^d}), \ \ \ F(r)=1+O(\frac{1}{r^{2d}}) ,\\  \label{GHCk}
k(r)&=& \frac{\sqrt{r^2-1}}{r}+\frac{\beta_{ n }}{r^d}+O(\frac{1}{r^{d+1}}).
\end{eqnarray}
Here 'e' denotes effective. Using the above solutions, we can work out
$ h^{(d)}_{\hat{i}\hat{j}}$ in the Fefferman-Graham
expansion. Interestingly, it takes exactly the same form as that of
Einstein gravity eq.(\ref{Einhij}), only replacing $M$ and $\beta_n$
by the effective counterparts  $M_e$ 
and $\beta_{ n}$. Comparing eqs.(\ref{hnCDholoTij}) with eq.
(\ref{hnCDfromTij}), we finally obtain
\begin{eqnarray}\label{GHChn}
&&\frac{h_n}{ C_T}=-2\pi n \frac{ M_e}{f_d},\\ \label{GHCCD}
&& \frac{C_D}{C_T}= \frac{d \pi ^2 n}{d+1}\big{[} (d-2) 
(\beta_{n}-\beta_{ 1})-\frac{M_e}{2} \big{]},
\end{eqnarray}
where 
$f_d$ and $C_T$ are given by \eq{fd} and \eq{CT}. 
$h_n$ and $C_D$ 
were first obtained for Einstein Gravity in
\cite{Balakrishnan:2016ttg} and for Gauss-Bonnet Gravity in
\cite{Dong123}. Here we derive them for the general higher curvature
gravity. 
It is remarkable that, 
when expressed in terms of $M_e$ and the $\b_{n}$'s,
the coefficients $h_n$ and $ C_D$ take on these very simple universal
forms \eq{GHChn}, \eq{GHCCD}. As a first check, our formulae agree with those of \cite{Balakrishnan:2016ttg,Dong123} for Einstein gravity and Gauss-Bonnet Gravity.
The holographic relations \eq{GHChn} and \eq{GHCCD}  are one of the main results 
we obtain for general dimensional CFTs. 
It should be mentioned that $G_N$ and $c^{(n)}_i$ appearing
in this section are actually  $ \tilde{G}_N$ and $ \tilde{c}^{(n)}_\ii$
defined in the action (\ref{GHCaction2}). For simplicity, we have
ignored the notation $\tilde{}\,$.

\subsection{CFTs in Three Dimensions}

In this section, we use the formulas obtained in the above section to
study the universal behaves of $h_n''(1)$ and $C_D''(1)$ for 3d
CFTs. We need to solve the E.O.M of general higher curvature to get 
the effective mass 
$M_e$ and $\beta_{ n}$. 
Note that the Gauss-Bonnet term is a total derivative in
four-dimensional spacetime. Without loss of generality, we can set $
c^{(2)}_1=0$. After some calculations, we derive
\begin{eqnarray}\label{3dHDfF}
f(r)&=&r^2-1+\frac{n-1}{r}-\frac{3 \left(r^5 (8 c^{(3)}_7-4
  c^{(3)}_8+5)+4 r^2 ( c^{(3)}_8-44  c^{(3)}_7)+144  c^{(3)}_7\right)
  (n-1)^2 }{8 r^6}\nonumber\\
&&+O(n-1)^3,\\
F(r)&=&1-\frac{9  (12 c^{(3)}_7+c^{(3)}_8)}{2 r^6}(n-1)^2+O(n-1)^2 .
\end{eqnarray}
One can see 
that 
these solutions obey the behaving (\ref{GHCfF}) and the effective mass
is given by
\begin{eqnarray}\label{3dHDMe}
M_e=-(n-1)+\frac{3 (8 c^{(3)}_7-4 c^{(3)}_8+5)}{8}(n-1)^2 +O(n-1)^2.
\end{eqnarray}
Note that we have used the conditions $f(r_H)=F(r_H)=0$ and
$T=\frac{1}{2\pi n}$ to fix the constants of integration for 
$f(r)$ and $F(r)$, with $r_H$ given by
\begin{eqnarray}\label{3dHDrh}
r_H=1-\frac{n-1}{2}+\frac{9}{16} (n-1)^2-\frac{9 c^{(3)}_7 (n-1)^2}{2}+O(n-1)^3.
\end{eqnarray}
Solving $k(r)$ up to order $O(n-1)^2$, we obtain
\begin{eqnarray}\label{3dHDk}
k(r)&=& \frac{\sqrt{r^2-1}}{r}+\frac{\beta_{
    n}}{r^3}+O(\frac{1}{r^{4}})\\ \label{3dHbeta}
\beta_{n}&=& \frac{n-1}{6}+\left(\frac{19 c^{(3)}_7}{2}-\frac{19
  c^{(3)}_8}{4}-\frac{41}{144}\right)(n-1)^2 +O(n-1)^3
\end{eqnarray}
Substituting eqs.(\ref{3dHDMe},\ref{3dHbeta}) into eqs.
(\ref{GHChn},\ref{GHCCD}), we obtain
\begin{eqnarray}\label{3dCHChn}
&&\frac{h_n}{ C_T}=\frac{1}{24} \pi ^3 (n-1)-\frac{ \pi ^3}{11520}
  (420+t_4)(n-1)^2+O(n-1)^3,\\ \label{3dGHCCD}
&& \frac{C_D}{C_T}=\frac{1}{2} \pi ^2 (n-1)-\frac{\pi ^2}{240}
  (100-t_4)(n-1)^2  +O(n-1)^3.
\end{eqnarray}
where we have used \cite{Sen:2014nfa} $t_4=720 (2 c^{(3)}_7-c^{(3)}_8)$. 

Now let us compare our holographic results with those of free
CFTs. $h_n$ for free fermions and free scalars are calculated in
\cite{Klebanov:2011uf,Lee:2014zaa,Dowker:2015qta,Dowker:2015pwa,Hung:2014npa}. And
it is proved in \cite{Dowker:2015qta,Dowker:2015pwa} that $C_D=d
\Gamma(\frac{d+1}{2})(\frac{2}{\sqrt{\pi}})^{d-1}h_n$ for free
fermions and scalars in three dimensions. For free Dirac fermions, we
have \cite{Hung:2014npa}
\begin{eqnarray}\label{Freefermions3d}
C_T=\frac{3}{16 \pi ^2},\; t_4=-4, \; h_n'(1)=\frac{\pi }{128},\; 
h_n''(1)=-\frac{13\pi }{960},\;  C_D'(1)=\frac{3}{32}, \; C_D''(1)=-\frac{13}{80},
\end{eqnarray}
which exactly match the holographic results
eqs.(\ref{3dCHChn},\ref{3dGHCCD}).  
However, similar to the case of 4d CFTs, mismatch appears for free
scalars. 
According to \cite{Lee:2014zaa,Dowker:2015qta}, it is 
\begin{eqnarray}\label{Freescalar3d}
C_T=\frac{3}{16 \pi ^2},\; t_4=4, \; h_n'(1)=\frac{\pi }{128},\; 
h_n''(1)=-\frac{17\pi }{960},\;  C_D'(1)=\frac{3}{32}, \; C_D''(1)=-\frac{17}{80},
\end{eqnarray}
for free complex scalars.
It is found in \cite{Lee:2014zaa,Hung:2014npa} there is discrepancy
for 
$ h_n''(1)$.
Here we note further that there is a  discrepancy in 
$C_D''(1)$ too. 
Similar to the 4d case, we find a combination of $ h_n''(1)$ and $C_D''(1)$,
\begin{eqnarray}\label{3duniversallaw}
\pi  C_D''(1)-16 h_n''(1)=\frac{\pi^3}{3} C_T(1+ \frac{t_4}{30})
\end{eqnarray}
which is obeyed by free scalars, free fermions and CFTs with gravity
dual.  In addition to free CFTs and
strongly coupled CFTs with gravity dual, it is interesting to
investigate whether 
the 
'universal law' \eq{3duniversallaw}
is  obeyed by more general CFTs.

\subsection{CFTs in Higher Dimensions}

Let us go on to discuss $h_n$ and $C_D$ in higher dimensions. Similar
to the cases of 3d CFTs and 4d CFTs, we need to solve the E.O.M in the
bulk to get the effective mass and $\beta_n$. Then we can derive $h_n$
and $C_D$ from the general formula eqs.(\ref{GHChn},\ref{GHCCD}).

By solving the E.O.M for the general higher curvature gravity
(\ref{GHCaction2}), we 
obtain 
\begin{eqnarray}\label{Gcasef}
f(r)&=&r^2-1-\frac{M_{\text{Ein}}}{r^{d-2}}+\frac{c^{(2)}_1
  f_1(r)+c^{(3)}_7 
f_7(r)+c^{(3)}_8 f_8(r)}{1+4(d-2)c^{(2)}_1}(n-1)^2+O(n-1)^3\nonumber\\
&=& r^2-1-\frac{M_{e}}{r^{d-2}}+O(\frac{1}{r^d}), \\  \label{GcaseF}
F(r)&=& 1+ \frac{c^{(3)}_7 F_7(r)+c^{(3)}_8
  F_8(r)}{1+4(d-2)c^{(2)}_1}(n-1)^2 
+O(n-1)^3=1+O(\frac{1}{r^{2d}}),\\ \label{Gcasek}
k(r)&=& \frac{\sqrt{r^2-1}}{r}+\frac{\beta_n}{r^d}+O(\frac{1}{r^{d+1}}),
\end{eqnarray}
where
\begin{eqnarray}\label{MEin}
M_{e}&=&-\frac{2}{d-1}(n-1)+\frac{(2d-3)(2d-1)}{(d-1)^3}(n-1)^2\nonumber\\
&&+ \frac{c^{(2)}_1 m_1+c^{(3)}_7 m_7+c^{(3)}_8
  m_8}{1+4(d-2)c^{(2)}_1}(n-1)^2+O(n-1)^3,
\eea 
and 
\bea \label{beta}
\beta_n&=&
\beta_1+\frac{1}{d(d-1)}(n-1)-\frac{4d^3-8d^2+d+2}{2d^2(d-1)^3}(n-1)^2\nonumber\\
&&+ \frac{c^{(2)}_1 b_1+c^{(3)}_7 b_7+c^{(3)}_8 b_8}{1+4(d-2)c^{(2)}_1}(n-1)^2+O(n-1)^3.
\eea
Here $f_1(r), f_7(r), f_8(r), F_7(r), F_8(r), m_1, m_2, m_3, b_1, b_7,
b_8 $ are determined by the E.O.M. We have 
worked out the solutions
case by case up to $d=9$.
Please refer to the appendix for these solutions.
In summary we obtain:
\begin{eqnarray}\label{generalCHChn}
&&\frac{h_n}{
    C_T}=2\pi^{\frac{d}{2}+1}\frac{\Gamma(\frac{d}{2})}{\Gamma(d+2)}
  (n-1)+\frac{h_n''(1)}{2C_T}(n-1)^2+O(n-1)^3,\\ 
\label{generalCHChn1}
&& \frac{C_D}{C_T}= \frac{2\pi^2}{d+1}
  (n-1)+\frac{C_D''(1)}{2C_T}(n-1)^2  +O(n-1)^3,
\end{eqnarray}
with $\frac{h_n''(1)}{C_T}$ and $\frac{C_D''(1)}{C_T}$ given by
\begin{eqnarray}\label{hngeneralD}
\frac{h_n''(1)}{C_T}&=&-\frac{2 \pi ^{\frac{d}{2}+1} \Gamma 
\left(\frac{d}{2}\right)}{(d-1)^3 d (d+1) \Gamma (d+3)} \big{[} d
  \left(2 d^5-9 d^3+2 d^2+7 d-2\right)
\nonumber\\
&+&(d-2) (d-3) (d+1) (d+2) (2 d-1) t_2+(d-2) \left(7 d^3-19 d^2-8
  d+8\right)t_4 \big{]},
\\ \label{CDgeneralD}
\frac{C_D''(1)}{C_T}&=&\frac{4 \pi ^2}{d+1}\big{[}
  \frac{1-d^2+d}{d^2-d}-\frac{(d-2)(d-3) }{(d-1)^2 d}\  t_2
  -\frac{(d-2) \left(3 d^2-7 d-8\right)}{(d-1)^2 d (d+1) (d+2)}\  t_4
  \big{]}.
\end{eqnarray} 
Note that the coefficients of $ t_2 $ and $t_4$ ($t_2$) in $h_n''(1)$
and $C_D''(1)$ (\ref{hngeneralD},\ref{CDgeneralD}) vanish when $d=2$
( $d=3$). This is the expected result, which can be regarded as a
check of our general formula (\ref{hngeneralD},\ref{CDgeneralD}). One
can also check that the general formulas
(\ref{hngeneralD},\ref{CDgeneralD}) reproduce the results of 3d and 4d
CFTs.
We remark that the holographic formula of $h_n''(1)$ (\ref{hngeneralD}) 
agrees with the those of \cite{Lee:2014zaa,Hung:2014npa}, which are
derived by using three-point functions of stress tensor. As they have
checked, 
the relation (\ref{hngeneralD}) for $h_n''(1)$
works well for free fermions
(up to $d=12$) but not for free scalars ($d>2$). 

Before we end this section, let us make some comments 
about the possible universal relation between $C_D$ and $h_n$.
For general dimensions, the generalization of the 4d
conjecture \eq{fbfc} is the statement \cite{Bianchi:2015liz}:
\be \label{CDhn-free}
C_D (n)=d
\Gamma(\frac{d+1}{2})(\frac{2}{\sqrt{\pi}})^{d-1}h_n (n).
\ee 
This relation can be motivated by the observation that
if one assume \eq{CDhn-free} holds for free
fermions and conformal tensor fields, one can prove $C_D''(1)$ of these 
fields exactly match the
holographic formula (\ref{CDgeneralD}). Turning the logic around, if
one assume free fermions and conformal tensor fields obey the 
holographic formulas
(\ref{hngeneralD},\ref{CDgeneralD}) \footnote{This is indeed the case
at least in three dimensions for free fermions.}, 
one can prove that 
the weaker relation
\be \label{Cnhn-free}
C_D''(1)=d\Gamma(\frac{d+1}{2})(\frac{2}{\sqrt{\pi}})^{d-1}h_n''(1)
\ee
holds 
in general dimensions. 
In proving these, we have found useful the relations \eq{CT}, \eq{t2},
\eq{t4} and that 
\begin{eqnarray}
&& A= \frac{1}{S_d^3}\big{[} \frac{d^3}{(d-1)^3}
    n_s-\frac{d^3}{d-3}\tilde{n}_t\big{]} ,\label{ABC1} \\
&& B=-\frac{1}{S_d^3}\big{[}
\frac{(d-2) d^3 }{(d-1)^3}n_s+ \frac{ d^2}{2}
\tilde{n}_f+\frac{(d-2)d^3}{d-3} \tilde{n}_t \big{]}, \label{ABC2}\\
&& C=-\frac{1}{S_d^3} \big{[} \frac{(d-2)^2 d^2 }{4
      (d-1)^3}n_s + \frac{d^2}{4} \tilde{n}_f +\frac{(d-2)d^3}{2(d-3)}
    \tilde{n}_t \big{]} , \label{ABC3}
\eea
where $S_d = 2 \pi^{d/2}/\Gamma(\frac{d}{2})$, $\tilde{n}_f =
 \tr(1)n_f =2^{[d/2]}n_f$, $n_f$ is the number of Dirac fermion, $\tr$
 is the Dirac trace and $ \tilde{n}_t$ denotes the number of degrees
 of freedom contributed by the $(n-1)$-form in even dimensions $d=2n$
 \cite{Buchel:2009sk}.  
However incompatiblity arises in the scalar sector as before.
Indeed using 
\eq{CT}, \eq{t2}, \eq{t4} and \eq{ABC1}-\eq{ABC3} in the
 holographic formulas  (\ref{hngeneralD},\ref{CDgeneralD}), we find
for a free theory with $n_s$ scalars,  
\begin{eqnarray}
C_D''(1)-d\Gamma(\frac{d+1}{2})(\frac{2}{\sqrt{\pi}})^{d-1}h_n''(1)
=\frac{(d-2)^4
  \pi ^{2-d}  \Gamma \left(\frac{d}{2}-1\right)^2}{16 (d-1)^3} n_s
\neq 0.
\eea
Thus problems only appear for scalars. This equation shows that
the relation \eq{Cnhn-free}
and the holographic formulas (\ref{hngeneralD},\ref{CDgeneralD}) 
cannot 
both be satisfied at the same time by free scalars.

That the relation \eq{CDhn-free} is  not compatible
with  the holographic results can also be seen from the consideration
of the positivity constraints \cite{Camanho:2009vw,Buchel:2009sk}
for  CFTs in general dimensions:
\begin{eqnarray}\label{energyconstraintscalar}
&&\text{Scalar Constraint}: \ \ 1+\frac{d-3 }{d-1}t_2+ \frac{d^2-d-4
  }{d^2-1}t_4  \ge 0,\\ \label{energyconstraintvector}
&&\text{Vector Constraint}: \ \ 1+\frac{d-3 }{2(d-1)}t_2- \frac{2
  }{d^2-1}t_4  \ge 0,\\ \label{energyconstrainttensor}
&&\text{Tensor Constraint}: \ \ 1-\frac{1 }{d-1}t_2- \frac{2 }
{d^2-1}t_4  \ge 0.
\end{eqnarray}
These constraints are consequences of the requirement of the
positivity of the energy fluxes.
Now it is easy to compute from 
(\ref{hngeneralD},\ref{CDgeneralD}) that 
\begin{eqnarray}\label{checkconjecture}
&&C_D''(1)-d\Gamma(\frac{d+1}{2})(\frac{2}{\sqrt{\pi}})^{d-1}h_n''(1)
\nonumber\\
&=&C_T\frac{2 \pi ^2 (d-2)}{(d-1)^2 d (d+1)} \left(1+\frac{d-3
  }{d-1}t_2+ \frac{d^2-d-4 }{d^2-1}t_4\right)  \ge 0,
\end{eqnarray}
where
in the last step we have used the unitarity constraint $C_T\ge 0$ and 
the  scalar constraint
\eq{energyconstraintscalar}.
This  shows that, unless 
$d=2$ or if the scalar constraint is saturated\footnote{The
  relation between
\eq{Cnhn-free}
  and lower bound of unitarity constraint (which is equivalent to the
  scalar constraint ) is observed for Gauss-Bonnet gravity for
  $d=4,5,6$ in \cite{Dong123}. Here we find this is a universal
  property for general higher curvature gravity in general
  dimensions.}, 
the relation
\eq{Cnhn-free}
and our holographic results \eq{hngeneralD}, \eq{CDgeneralD}
cannot both be satisfied at the same time.

All in all, it is therefore 
interesting to look for a different relation between 
$C_D''(1)$ and $h_n''(1)$ like those of \eq{Conjecture} for the 4d
case 
and \eq{3duniversallawNew} for the 3d case, 
that would hold for all free theories as well as strongly coupled dual
theories. 
To do so, we need the information of  $h_n''(1)$ and $C_D''(1)$ of
free scalars. 
$h_n''(1)$ of free scalars is discussed in
\cite{Casini:2010kt,Dowker:2015qta,Hung:2014npa}  in general
dimensions.
However, so far we do not know $C_D''(1)$ in dimensions higher than
four ($d> 4$). 
On the other hand, if we assume
\eq{Cnhn-free}
holds for free scalars in general dimensions 
as has been suggested in 
\cite{Bianchi:2015liz},
then
we obtain 
\begin{eqnarray}\label{universallawinGD}
&&C_D''(1)-2(d-1)\Gamma(\frac{d+1}{2})(\frac{2}{\sqrt{\pi}})^{d-1}
h_n''(1)\nonumber\\
&=&\frac{4 \pi ^2 (d-2)}{d (d+1)} C_T\big{[} 1+\frac{d-3}{d(d-1) } t_2
+\frac{4 \left(d^2-2 d-2\right)}{(d-1) d (d+1) (d+2)} t_4 \big{]},
\end{eqnarray}
which is such a  'universal law' obeyed by free scalars, free
fermions, free conformal tensor fields and CFTs with holographic
dual. 
Please refer to the appendix for the derivation of
eq.(\ref{universallawinGD}). As a quick check,
eq.(\ref{universallawinGD}) reproduces \eq{Conjecture} and
\eq{3duniversallawNew} for 4d and 3d CFTs, respectively. 
It is interesting
to find out if the 'universal law' (\ref{universallawinGD})
is indeed valid 
for  general CFTs. We leave this interesting problem 
and related questions 
for future work. 

In summary, our holographic results
(\ref{hngeneralD},\ref{CDgeneralD}) are obeyed by free fermions and
conformal tensors but are violated by free scalars. According to
\cite{Bianchi:2015liz}, it seems that the free CFTs satisfy
\eq{Cnhn-free}.
However,
as we have proven above, this relation does not agree with
eqs.(\ref{hngeneralD},\ref{CDgeneralD}). So neither the relation
\eq{Cnhn-free} nor the holographic relations \eq{hngeneralD},
\eq{CDgeneralD}) can be universally true for all CFTs.
Instead, we
find 
that the
suitably combined relation \eq{universallawinGD} is satisfied by 
free CFTs (including scalars) as well as by CFT with holographic
duals, and stands a 
chance to be a universal relation satisfied by all CFTs.

\section{Conclusions}

In this paper, we 
have investigated 
the universal terms of holographic
R\'enyi entropy for 4d CFTs. Universal relations between the
coefficients $f_a''(1), f_b'(1), f_c'(1)$ in the logarithmic terms of
R\'enyi entropy and the parameters $c, t_2, t_4$ of stress tensor
two-point and three-point functions are found. Interestingly, these
relations are also obeyed by weakly coupled CFTs such as free fermions
and vectors but are violated by scalars. Similar to the case of
$f_a''(1)$ \cite{Lee:2014zaa}, one expects that the discrepancy for
scalars comes from the boundary contributions to the modular
Hamiltonian. Remarkably, 
We have found that
there is a combination of our holographic
relations which 
is satisfied by all the free CFTs including
scalars. We conjecture 
that this combined relation 
\eq{ConjectureNew}
is universal for general
CFTs in four dimensional spacetime.
For general spacetime dimensions,
we obtain the holographic
dual of $h_n$ and $C_D$ for general higher curvature gravity.
Our holographic results together with the positivity of
energy flux imply $C_D''(1)\ge
d\Gamma(\frac{d+1}{2})(\frac{2}{\sqrt{\pi}})^{d-1}h_n''(1)$. And the
equality is satisfied by free fermions and the conformal tensor fields
if they obey the holographic universal laws. 
We also find
there are similar holographic universal laws of 
$h_n''(1)$ and $C_D''(1)$. By assuming \eq{Cnhn-free}
for free CFTs, 
we find 
that for general dimensions, the relation \eq{universallawinGD}
is obeyed by all the free CFTs as well as by CFTs with holographic
duals.
It is interesting to test these 'universal laws'
by studying more general CFTs. We leave a careful study of this
problem to future work. 

\section*{Acknowledgements}

R. X. Miao thank Yau Mathematical Sciences Center for hospitality
during the early stages of this work. In particular, R. X. Miao wish
to thank Prof. W. Song, Q. Wen and J. F. X for helpful discussions and
kind help during the stay at YMSC. This work is supported in part by
the National Center of Theoretical Science (NCTS) and the grant MOST
105-2811-M-007-021 of the Ministry of Science and Technology of
Taiwan.

 \appendix

\section{Equivalence between two Stress Tensors}

In the analysis in the main text, we have considered in section 2
the Brown-York boundary stress
tensor  eq.(\ref{GHCTij}) 
in section 2, and in section 4  
the holographic stress tensor  eq.(\ref{hnCDholoTij}). 
As we have mentioned in section 4, they are actually equivalent up to a
rescaling and some functions of $g_{(0) ij}$ \cite{deHaro:2000vlm} 
that are irrelevant:
\begin{eqnarray}\label{relation1}
<T_{ij}>=\lim_{z \to 0} \frac{1}{z^{d-2}} T_{\partial M ij}.
\end{eqnarray}
Here the LHS is the holographic stress tensor and the RHS is the
Brown-York boundary stress tensor.
In this appendix, we shall prove that, by applying the 
stress tensor eq.(\ref{hnCDholoTij}) instead of eq.(\ref{GHCTij}) in
the approach of section 2, we obtain the same results for $f_b(n)$ and
$f_c(n)$. This is can be regarded as a double check of our results.

The key point in section 2 is that the change in the partition function
is govern by the stress tensor one-point
function\begin{eqnarray}\label{changeZ}
\delta\log Z_n= \frac{1}{2}\int_{\partial M}dx^4 \sqrt{\gamma}
T_{\partial M}^{\ \ ij} \delta \gamma_{ij}
\end{eqnarray}
From eq.(\ref{relation1}) and the asymptotic AdS metric in the FG gauge
eq.(\ref{hnCDFGAdS}), one can rewrite it in terms of $<T_{ij}>$ and
$\delta g_{(0)ij}$ as
\begin{eqnarray}\label{changeZ0}
\delta\log Z_n= \frac{1}{2}\int dx^4 \sqrt{g_{(0)}}< T^{ij}> \delta g_{(0)ij}
\end{eqnarray}
The boundary metric $ g_{(0)ij}$ is given by (2.18) of [10]
\begin{eqnarray}\label{metric}
ds^2=d\tau^2+\frac{1}{\rho^2}\big{(} d\rho^2+
[\delta_{\hat{i}\hat{j}}+
2 \bar{K}^a_{\hat{i}\hat{j}}x_a+Q^{ab}_{\hat{i}\hat{j}}x_ax_b]
dy^{\hat{i}} dy^{\hat{j}} \big{)}+O(K^2).
\end{eqnarray}
Actually, we can ignore the $Q$ terms above, since it is of order $O(K^2)$.
For simplicity, we focus on the case of traceless extrinsic 
curvature $K^{a\hat{ i}}_{\ \ \hat{i}}=0$ as in sect.2. Using
eqs.(\ref{hnCDholoTij},\ref{Einhij}),  we can derive the stress tensor
in $\hat{i}\hat{j}$ components for 4d CFTs as
\begin{eqnarray}\label{Tij}
<T_{\hat{i}\hat{j}}> =\frac{4}{f_4}C_T
h^{(4)}_{\hat{i}\hat{j}}=\frac{4}{f_4}C_T\frac{1}{\rho^2}\big{[}
  (\frac{1}{4}M_e+ g_0)  \delta_{\hat{i}\hat{j}}+
  (\frac{1}{2}M_e+2\beta_{n}+ k_0) \bar{K}^a_{\hat{i}\hat{j}} x_a
  \big{]}+O(K^2).
\end{eqnarray}
From the above two equations, we get
\begin{eqnarray}\label{Tuiuj}
<T^{\hat{i}\hat{j}} >=\frac{4}{f_4}C_T
h^{(4)}{}^{\hat{i}\hat{j}}=\frac{4}{f_4}C_T\rho^2\big{[}
  (\frac{1}{4}M_e+ g_0)  \delta^{\hat{i}\hat{j}}+
  (-\frac{1}{2}M_e+2\beta_{n}+ k_0-4g_0) \bar{K}^{a \hat{i}\hat{j}}
  x_a   \big{]}+O(K^2).
\end{eqnarray}

Substituting eq.(\ref{Tuiuj}) and $ \delta
g_{(0)\hat{i}\hat{j}}=\frac{1}{\rho^2}(2 
\delta\bar{K}^a_{\hat{i}\hat{j}}x_a+  \delta
Q^{ab}_{\hat{i}\hat{j}}x_ax_b)$ 
into eq.(\ref{changeZ0}), we get
\begin{eqnarray}\label{changeZ10}
\delta\log Z_n= \frac{1}{2}\int_{\partial M}dx^4 \sqrt{g_{0}}
T^{\hat{i}\hat{j}} \delta g_{(0)\hat{i}\hat{j}}
\end{eqnarray}
Integrating eq.(\ref{changeZ10}) and selecting the logarithmic
divergent terms, we obtain
\begin{eqnarray}\label{changeZ1}
\log Z_n=-\log \epsilon \int_{\Sigma} dy^2 \frac{2n \pi}{f_4}C_T [
  3(\frac{1}{4}M_e+ g_0)  C^{ab}_{\ \ ab} +
  (-\frac{1}{2}M_e+2\beta_{n}+ k_0-4g_0) \tr \bar{K}^2  ]
\end{eqnarray}
where we have used $ C^{ab}_{\ \ ab} = \frac{1}{3}Q_a^{a
  \hat{i}}{}_{\hat{i}}$ in the above derivations. Using
eq.(\ref{changeZ1}) and $M_e(1)=0$, we obtain the logarithmic
divergent terms of R\'enyi entropy 
\begin{eqnarray}\label{renyientropy}
S_n&=&\frac{\log Z_n- n \log Z_1}{1-n}\nonumber\\
&=&\log \epsilon  \frac{n}{n-1}\frac{\pi}{f_4}C_T \int_{\Sigma} dy^2
[ \frac{3}{2}M_e  C^{ab}_{\ \ ab} +
  (- M_e+4(\beta_{n}-\beta_1)) \tr \bar{K}^2
] \label{renyientropy1}\\
&=&\log \epsilon \int_{\Sigma} dy^2  [\frac{f_b(n)}{2\pi} \tr \bar{K}^2
  -\frac{ f_c(n)}{2\pi}  C^{ab}_{\ \ ab} ].\label{renyientropy2}
\end{eqnarray}
Notice that the constants $g_0$ and $k_0$ are canceled automatically
in the above calcultions.
Comparing eq.(\ref{renyientropy1}) and eq.(\ref{renyientropy2}) and
using  $C_T/f_4=\frac{c}{2\pi^2}$, we finally obtain
\begin{eqnarray}\label{fbapp}
f_b(n)&=& \frac{n(4\beta_n-4\beta_1-M_e)}{n-1}c,\\
f_c(n)&=& -\frac{3 n M_e}{2(n-1)}c.\label{fcapp}
\end{eqnarray}

Recall that $M_e$ and $\beta_n$ are given by
\begin{eqnarray}\label{effectivemassapp}
M_e&=&-\frac{2}{3}(n-1)+\frac{ (336c^{(2)}_1+192
c^{(3)}_7-96c^{(3)}_8+35)}{27(1+8 c^{(2)}_1)} (n-1)^2+ O(n-1)^3\\
\beta_n&=&-\frac{1}{8}+\frac{n-1}{12}+\frac{ (-600 c^{(2)}_1+4224
  c^{(3)}_7-2112  c^{(3)}_8-67)}{432
  (1+8c^{(2)}_1)}(n-1)^2+O(n-1)^3.\label{effectivebapp}
\end{eqnarray}
Substituting eqs.(\ref{effectivemassapp},\ref{effectivebapp}) and $
c=\frac{\pi}{8G_N}(1+8c^{(2)}_1)$ into eqs.(\ref{fbapp},\ref{fcapp}),
we reproduce the results \eq{GHCfc1} and \eq{GHCfb1} in sect. 2. So
the stress tensor eq.(\ref{hnCDholoTij}) indeed yields the same
results for 4d CFTs as the stress tensor eq.(\ref{GHCTij}).

Substituting $h_n/(n-1)=\frac{2}{3\pi}f_c(n)$,
$C_D/(n-1)=\frac{16}{\pi^2}f_b(n)$ and $c=\pi^4 C_T/40$ [12] into
eqs.(\ref{fbapp},\ref{fcapp}), one can also reproduce $h_n$ and $C_D$
eqs.(\ref{GHChn},\ref{GHCCD}) for 4d CFTs in section 4.

\section{Solutions in General Higher Curvature Gravity}

In this appendix we provide the solutions to E.O.M of the general
higher curvature gravity (\ref{GHCaction2}), which are found to be
useful for the derivations of holographic $h_n$ and $C_D$ in
sect.4.2. For simplicity, we work in the perturbative framework of the
coupling constants $c^{(n)}_i$. To derive $h_n''(1)$ and $C_D''(1)$ in
terms of $C_T, t_2, t_4$, we can further set $c^{(2)}_1=0$ in
dimensions except $d=4$. The solutions for $d=3$ and $d=4$ are given
in sect. 4.1 and sect.2.2. Below we list the key results for
$d=5,6,7,8,9$ up to $O(n-1)^2$.

5d CFTs:
\begin{eqnarray}
f(r)&=&r^2-1+\frac{n-1}{2 r^3}\nonumber\\
&&-\frac{3  \left(3 r^7 (104 c^{(3)}_7-34  c^{(3)}_8+7)-8 r^2 (1044
  c^{(3)}_7-9  c^{(3)}_8)+7200 c^{(3)}_7\right)}
{64 r^{10}} (n-1)^2
+O(n-1)^3,\nonumber\\
F(r)&=&1-\frac{45  (22  c^{(3)}_7+ c^{(3)}_8)}{4 r^{10}} (n-1)^2 +O(n-1)^3,\\
k(r)&=& \frac{\sqrt{r^2-1}}{r}+\frac{1}{r^5}\frac{(n-1) (160-(n-1)
  (-30840 c^{(3)}_7+15990
  c^{(3)}_8+307))}{3200}+O(\frac{1}{r^{d+1}}).\nonumber
\eea

6d CFTs:
\begin{eqnarray}
f(r)&=&r^2-1+\frac{2 (n-1)}{5 r^4}\nonumber\\
&&-\frac{3 \left(r^8 (992 c^{(3)}_7-232c^{(3)}_8+33)-80 r^2 (328
  c^{(3)}_7-2 c^{(3)}_8)+23040c^{(3)}_7\right)}{125
  r^{12}} (n-1)^2  +O(n-1)^3,\nonumber\\
F(r)&=&1-\frac{72  (132 c^{(3)}_7+5 c^{(3)}_8)}{25 r^{12}} (n-1)^2 +O(n-1)^3,\\
k(r)&=& \frac{\sqrt{r^2-1}}{r}+\frac{1}{r^6}\frac{(n-1) (75-2 (n-1)
  (-10548 c^{(3)}_7+5733
  c^{(3)}_8+73))}{2250}+O(\frac{1}{r^{d+1}}).\nonumber
\eea

7d CFTs:
\begin{eqnarray}
f(r)&=&r^2-1+\frac{n-1}{3 r^5}\nonumber\\
&&+\frac{\left(r^9 (-7320 c^{(3)}_7+1320 c^{(3)}_8-143)+900 r^2 (220
  c^{(3)}_7-c^{(3)}_8)-176400c^{(3)}_7\right)}{216
  r^{14}}(n-1)^2 +O(n-1)^3,\nonumber\\
F(r)&=&1-\frac{35 (92 c^{(3)}_7+3 c^{(3)}_8)}{6 r^{14}}(n-1)^2 +O(n-1)^3,\\
k(r)&=& \frac{\sqrt{r^2-1}}{r}+\frac{1}{r^7}\frac{(n-1) ((n-1) (192696
  c^{(3)}_7-109704
  c^{(3)}_8-989)+504)}{21168}+O(\frac{1}{r^{d+1}}).\nonumber
\eea

 8d CFTs:
\begin{eqnarray}
f(r)&=&r^2-1+\frac{2 (n-1)}{7 r^6}\nonumber\\
&&-\frac{3 \left(r^{10} (5088 c^{(3)}_7-744c^{(3)}_8+65)-504 r^2
  (284c^{(3)}_7-c^{(3)}_8)+129024 c^{(3)}_7\right)}{343
  r^{16}} (n-1)^2 +O(n-1)^3,\nonumber\\
F(r)&=&1-\frac{144  (244 c^{(3)}_7+7c^{(3)}_8)}{49 r^{16}}(n-1)^2 +O(n-1)^3,\\
k(r)&=& \frac{\sqrt{r^2-1}}{r}+\frac{1}{r^8}\frac{(n-1) ((n-1) (194304
  c^{(3)}_7-115392c^{(3)}_8-773)+392)}{21952}+O(\frac{1}{r^{d+1}}).\nonumber
\eea

9d CFTs:
\begin{eqnarray}
f(r)&=&r^2-1+\frac{n-1}{4 r^7}\nonumber\\
&&-\frac{3 \left(r^{11} (9464 c^{(3)}_7-1162 c^{(3)}_8+85)-784 r^2
  (356 c^{(3)}_7-c^{(3)}_8)+254016 c^{(3)}_7\right)}{512
  r^{18}} (n-1)^2 +O(n-1)^3,\nonumber\\
F(r)&=&1-\frac{189  (39 c^{(3)}_7+c^{(3)}_8)}{8 r^{18}} (n-1)^2 +O(n-1)^3,\\
k(r)&=& \frac{\sqrt{r^2-1}}{r}+\frac{1}{r^9}\frac{(n-1) ((n-1) (715608
  c^{(3)}_7-441234
  c^{(3)}_8-2279)+1152)}{82944}+O(\frac{1}{r^{d+1}}).\nonumber
\eea

Using these solutions, we can derive $M_e$ and $\beta_n$ from eqs.(\ref{GHCfF},\ref{GHCk}) as
\begin{eqnarray}\label{Megeneral}
M_{e}&=&-\frac{2}{d-1}(n-1)+\frac{(2d-3)(2d-1)}{(d-1)^3}(n-1)^2\nonumber\\
&&+ \frac{12 (d-2) \left(d^3-6 d^2+11 d-4\right)}{(d-1)^3} c^{(3)}_7 (n-1)^2\nonumber\\
&&-\frac{3 (d-2) \left(3 d^2-9 d+4\right)}{(d-1)^3} c^{(3)}_8 (n-1)^2+O(n-1)^3,
\eea 
and 
\bea \label{betageneral}
\beta_n&=&
\beta_1+\frac{1}{d(d-1)}(n-1)-\frac{4d^3-8d^2+d+2}{2d^2(d-1)^3}(n-1)^2\nonumber\\
&&+\frac{6 \left(d^4+2 d^3-21 d^2+36 d-16\right)}{(d-1)^3 d} c^{(3)}_7 (n-1)^2\nonumber\\
&&-\frac{3 \left(4 d^4-17 d^3+35 d^2-40 d+16\right)}{2 (d-1)^3 d} c^{(3)}_8 (n-1)^2+O(n-1)^3.
\eea
Recall that we have
\bea \label{c7}
c^{(3)}_7 &=&\frac{2 \left(d^2+3 d+2\right) t_2+(7 d+4) t_4}{12 (d-1) d \left(d^3-d^2-10 d-8\right)},\\ \label{c8}
c^{(3)}_8 &=&\frac{\left(d^2+3 d+2\right) t_2+(3 d+4) t_4}{3 d \left(d^4-2 d^3-9 d^2+2 d+8\right)}.
\eea
Substituting eqs.(\ref{Megeneral}-\ref{c8})  into the holographic formula (\ref{GHChn},\ref{GHCCD}), we can derive $h_n$ and $C_D$ eqs.(\ref{generalCHChn}-\ref{GHCCD}) in sect. 4.2. 

\section{Universal Laws in General Dimensions}

$h_n$ for free comformally coupled scalars in even-dimensional space-time are calculated in \cite{Hung:2014npa}
\bea \label{hnfreescalar}
h_n=\frac{(2\pi)^{1-d}}{d-1} \sum_{j=0}^{(d-4)/2}a_{j, l-1}^{(0)}(2j-d+1)\pi^{2j-d/2}(n^{2j-d+1}-n)\Gamma(\frac{d}{2}-j)\zeta(d-2j),
\eea
where $d=2l+2$ and $a_{j, l-1}^{(0)}$ are defined by
\bea \label{ajl}
P_{l-1}^{(0)}(t)&=& \sum_{j=0}^{l-1}a_{j, l-1}^{(0)} t^j\nonumber\\
&=&\lim_{\rho\to 0} (4\pi t)^{l}\big{(} \frac{-1}{2\pi \sinh \rho} \frac{\partial}{\partial \rho}\big{)}^l \exp(-\frac{\rho^2}{4t}).
\eea
The first few polynomials are given by
\bea \label{Plscalar}
P_{0}^{(0)}(t)&=&1,\nonumber\\
P_{1}^{(0)}(t)&=&1+\frac{2 t}{3},\nonumber\\
P_{2}^{(0)}(t)&=&1+2t +\frac{16 t^2}{15},\nonumber\\
P_{3}^{(0)}(t)&=&1+4t+\frac{28 t^2}{5}+\frac{96 t^3}{35} ,\nonumber\\
P_{4}^{(0)}(t)&=&1+\frac{20 t}{3}+\frac{52 t^2}{3}+\frac{1312 t^3}{63}+\frac{1024 t^4}{105} ,\nonumber\\
P_{5}^{(0)}(t)&=&1+10 t+\frac{124 t^2}{3}+\frac{5560 t^3}{63}+\frac{30656 t^4}{315} +\frac{10240 t^5}{231},\nonumber\\
P_{6}^{(0)}(t)&=&1+14 t +84 t^2+\frac{2488 t^3}{9}+\frac{4736 t^4}{9}+\frac{30208 t^5}{55}+\frac{245760 t^6}{1001}.
\eea
From eq.(\ref{hnfreescalar}), it is easy to derive $h_n''(1)$ as
\bea \label{bbhnfreescalar}
h_n''(1)=\frac{(2\pi)^{1-d}}{d-1} \sum_{j=0}^{(d-4)/2}a_{j, l-1}^{(0)}(2j-d+1)^2(2j-d)\pi^{2j-d/2}\Gamma(\frac{d}{2}-j)\zeta(d-2j),
\eea
Unfortunately, now we do not have a formula of $C_D(n)$. It seems that  $C_D''(1)=d\Gamma(\frac{d+1}{2})(\frac{2}{\sqrt{\pi}})^{d-1}h_n''(1)$ holds for free scalars in general dimensions \cite{Bianchi:2015liz}. This is at least the case for $d=3$ and $d=4$. With this assumption, now we are ready to derive the 'universal law' (\ref{universallawinGD}). 

We require that the 'universal laws' are obeyed by both the free scalars and the holographic CFTs. From eqs.(\ref{bbhnfreescalar},\ref{generalCHChn},\ref{CDgeneralD}) and the assumption mentioned above, we finially obtain  the 'universal law' (\ref{universallawinGD}) in general dimensions. In the derivations, we have used the following useful formula
\bea \label{GULawuesfulformula}
 \sum_{j=0}^{(d-4)/2}a_{j, l-1}^{(0)}\frac{4 d (d+2) (d-2 j-1)^2 \pi ^{-d+2 j-\frac{1}{2}} \Gamma \left(\frac{d+3}{2}\right) \zeta (d-2 j) \Gamma \left(\frac{d}{2}-j+1\right)}{(d (3 d-2)-4) \Gamma \left(\frac{d}{2}\right)^2}=1.
\eea
Using eq.(\ref{Plscalar}), we verified that this identity holds up to $d=16$.

It should be mentioned that although we focus on even dimensions in the above discussions. We have checked that  the 'universal law' (\ref{universallawinGD}) produces correct result (\ref{3duniversallawNew}) in three dimensions. So it is expected that  eq.(\ref{universallawinGD}) works well in general odd dimensions too.

\end{document}